\documentclass[prb,twocolumn,amsmath]{revtex4-1}

\usepackage{graphicx}

\usepackage{amssymb,amsfonts,amsmath}

\def \>{\rangle} 
\def \<{\langle} 
 
\def\be{\begin{equation}} 
\def\ee{\end{equation}} 
\def\longrightharpoonup{\relbar\joinrel\rightharpoonup}
\def\longleftharpoondown{\leftharpoondown\joinrel\relbar}

\def\longrightleftharpoons{
  \mathop{
    \vcenter{
      \hbox{
      \ooalign{
        \raise1pt\hbox{$\longrightharpoonup\joinrel$}\crcr
	  \lower1pt\hbox{$\longleftharpoondown\joinrel$}
	  }
      }
    }
  }
}

\newcommand \bea {\begin{eqnarray}} 
\newcommand \eea {\end{eqnarray}}

\begin{document}

\title{Landauer in the age of synthetic biology: energy consumption and information processing in biochemical networks}

\author{Pankaj Mehta}
\affiliation{Dept. of Physics, Boston University, Boston, MA 02215}
\date{\today}

\author{Alex H. Lang}
\affiliation{Dept. of Physics, Boston University, Boston, MA 02215}
\date{\today}

\author{David J. Schwab }
\affiliation{Dept. of Physics, Northwestern, Evanston, IL}
\date{\today}
\begin{abstract}

A central goal of synthetic biology is to design sophisticated synthetic cellular circuits that can perform complex computations and information processing tasks in response to specific inputs. The tremendous advances in our ability to understand and manipulate cellular information processing networks raises several fundamental physics questions: How do the molecular components of cellular circuits exploit energy consumption to improve information processing? Can one utilize ideas from thermodynamics to improve the design of synthetic cellular circuits and modules?  Here, we summarize recent theoretical work addressing these questions. Energy consumption in cellular circuits serves five basic purposes: (1) increasing specificity, (2) manipulating dynamics,  (3) reducing variability, (4) amplifying signal, and (5) erasing memory. We demonstrate these ideas using several simple examples and discuss the implications of these theoretical ideas for the emerging field of synthetic biology. We conclude by discussing how it may be possible to overcome these limitations using ``post-translational"  synthetic biology that exploits reversible protein modification.
\end{abstract}

\maketitle

Cells live in complex and dynamic environments. They sense and respond to both external environmental cues and to each other through cell-to-cell communication. Adapting to changing environments often requires cells to perform complex information processing, and cells have developed elaborate signaling networks to accomplish this feat. These biochemical networks are ubiquitous in biology, ranging from the quorum-sensing\cite{Miller2001Quorum} and chemotaxis networks\cite{Sourjik2012Responding} in single-celled organisms to developmental networks in higher organisms\cite{Logan2004The-WNT-Signaling}. Inspired by both these natural circuits and physical computing devices, synthetic biologists are designing sophisticated synthetic circuits that can perform complicated ``computing-like'' behaviors. Synthetic biologists have designed gene circuits executing a wide range of functionalities including switches\cite{Gardner2000Construction}, oscillators\cite{Elowitz2000A-synthetic}, counters\cite{Friedland2009Synthetic}, and even cell-to-cell communicators\cite{Danino2010A-synchronized}.  

Despite these successes, many challenges to harnessing the full potential of synthetic biology persist\cite{Mukherji2009Synthetic,Purnick2009The-second,Khalil2010Synthetic,Nandagopal2011Synthetic,Ruder2011Synthetic,Keung2015Chromatin,Cameron2014A-brief,Cheng2012Synthetic}. While there are guiding principles to synthetic biology\cite{Way2014Integrating}, actual construction of synthetic circuits often proceeds in an ad-hoc manner through a mixture of biological intuition and trial-and-error. Furthermore, the functionality and applicability is limited by a dearth of biological components\cite{Kwok2010Five}. For this reason, it would be helpful to identify general principles that can improve the design of synthetic circuits and help guide the search for new biological parts.  One promising direction along these lines is recent work examining  the relationship between the information processing capabilities of these biochemical networks and their energetic costs (technically this is usually a cost in free energy, but for the sake of brevity we will refer to this as energy). Energetic costs place important constraints on the design of physical computing devices\cite{Landauer1961Irreversibility} as well as on neural computing architectures in the brain and retina\cite{Laughlin2001Energy}, suggesting that thermodynamics may also influence the design of cellular information processing networks. As the field of synthetic biology seeks to assemble increasingly complex biochemical networks that exhibit robust, predictable behaviors, natural questions emerge: What are the physical limitations (thermodynamic and kinetic) on the behavior and design of these biological networks? How can one use energy consumption to improve the design of synthetic circuits?

In a classic paper written at the advent of modern computing\cite{Landauer1961Irreversibility}, Landauer asked analogous questions about physical computing devices. He argued that a central component of any general purpose computing device is a memory module that can be ``reset'' to a predefined state, and pointed out that such a device must obey certain thermodynamic and kinetic constraints. In particular, he convincingly argued that resetting memory necessarily leads to power dissipation, implying that heat generation and energy consumption are unavoidable consequences of the computing process itself. The paper also outlined three general sources of error resulting from kinetic and thermodynamic considerations: incomplete switching between memory states due to long switching times, the decay of stored information due to spontaneous switching, and what he called a ``Boltzmann'' error due to limited energy supplies. Furthermore, the paper showed that there exist fundamental trade-offs between these types of errors and energetic costs in these 
memory devices. These considerations suggested general strategies for designing new devices and parts for physical memory modules.

The goal of this review is to synthesize recent theoretical work on thermodynamics and energy consumption in biochemical networks and discuss the implications of this work for synthetic biology. Theoretical papers in this field are often highly technical and draw on new results in non-equilibrium statistical mechanics. For this reason, our goal is to organize the insights contained in these papers\cite{Qian2005Thermodynamics,Qian2007Phosphorylation,Zhang2012Stochastic,Ge2012Stochastic,Barato2013Information-theoretic,Lan2012The-energy-speed-accuracy,Govern2012Fundamental,Sagawa2012Nonequilibrium,Still2012Thermodynamics,Barato2013Information,Govern2013How-biochemical,Becker2013Prediction,Bo2015Thermodynamic,Govern2014Energy,Govern2014Optimal,Kaizu2014Berg,Iyengar2014A-cellular,Selimkhanov2014Accurate,Murugan2014Discriminatory,Barato2014Efficiency,Ito2014Maxwells,Sartori2014Thermodynamic,Bo2015Thermodynamic,Ouldridge2015On-the-connection,Hartich2015Nonequilibrium} into a few simple, broadly applicable principles.  We find that energy consumption in cellular circuits tends to serve five basic purposes: (1) increasing specificity, (2) manipulating dynamics,  (3) reducing variability, (4) amplifying signal, and (5) erasing memory. Furthermore, for each of these categories, there exist implicit tradeoffs between power consumption and dynamics.

In the future, energetic costs are likely to become an increasingly important consideration in the design of synthetic circuits. Presently, synthetic biology is adept at making circuits that can be controlled and manipulated by  external users  by, for example, adding or removing small signaling molecules. A major challenge facing the field is to move beyond such externally controlled circuits to autonomous circuits that can function in diverse environments for extended periods of time. Such autonomous circuits must be able to accurately sense the external environment, amplify small signals, and store information -- processes that require or can be improved through energy consumption. Energy consumption necessarily imposes a fitness cost on cells harboring the synthetic circuit, and over many generations, even a small fitness cost can cause synthetic circuits to be lost due to competition. For this reason, understanding how the information processing capabilities of a biochemical network are related to its energy consumption is an important theoretical problem in synthetic biology.

Beyond synthetic biology, biochemical networks offer a unique setting to explore fundamental physics questions in non-equilibrium statistical mechanics. Recently there has been a surge of interest among physicists in the relationship between information and thermodynamics\cite{Bennett1982The-thermodynamics,Bennett2003Notes}. For example, using sophisticated optical traps groups have recently experimentally tested LandauerÕs principle\cite{Berut2012Experimental,Jun2014High-Precision}, and there is an active debate on how to extend LandauerÕs principle to quantum regimes\cite{Vedral2014Quantum}. A flurry of recent work has focused on extending concepts like entropy and free-energy to non-equilibrium regimes, often using information theoretic concepts\cite{Vaikuntanathan2011Modeling,Jarzynski2011Equalities,Mandal2012Work,Vaikuntanathan2014Dynamic,Diamantini2014Generalized,Das2014Capturing,Parrondo2015Thermodynamics}. Living systems are perhaps the most interesting example of non-equilibrium systems, and thinking about information and thermodynamics in the context of cells is likely to yield new general insights into non-equilibrium physics.

\section{Increasing Specificity}
One common role of energy consumption in biochemical circuits is to increase the specificity of an enzyme or signaling pathway. The most famous example of this is kinetic proofreading. In a landmark paper\cite{Hopfield1974Kinetic}, John Hopfield showed how it is possible to increase the specificity of an enzyme beyond what would be expected from equilibrium thermodynamics by consuming energy and driving the system out of equilibrium. Kinetic proofreading-type mechanisms are also thought to underlie the exquisite specificity of eukaryotic pathways such as the TCR signaling network \cite{Mckeithan1995Kinetic}, in which a few-fold difference in the affinities between molecules can lead to several orders of magnitude difference in response. A full review of kinetic proofreading and all its applications is beyond the scope of this review, but we highlight some important lessons for synthetic biology. 

The first general principle that emerges from kinetic proofreading is that greater specificity requires greater energy consumption. In particular, the error rate in kinetic proofreading depends exponentially on the amount of energy consumed in each step of the proofreading cascade. This increased specificity comes at the expense of a more sluggish dynamic response (see \cite{Murugan2012Speed,Murugan2014Discriminatory} for an interesting exploration of this tradeoff). This highlights a second theme about energy consumption: there generally exist trade-offs between greater specificity and other desirable properties such as a fast dynamical response or sensitivity to small signals. 

The latter trade-off is clearest in the context of non-specific activation of an output in a synthetic circuit. For example, in a transcriptional synthetic circuit an output protein may be produced at low levels even in the absence of an input signal. A common strategy for dealing with such background levels of activation is to place a strong degradation tag on the protein that increases its degradation rate\cite{Cameron2014Tunable}. This ensures that in the absence of an activating signal, proteins are quickly degraded. However, increasing the degradation rate clearly comes at a steep energetic cost as more proteins have to be produced to reach the same steady-state. At the same time, the gene circuit loses sensitivity to small input signals due to their fast degradation.

\section{Manipulating Dynamics}

\begin{figure}[t]
\includegraphics[width=0.5\textwidth]{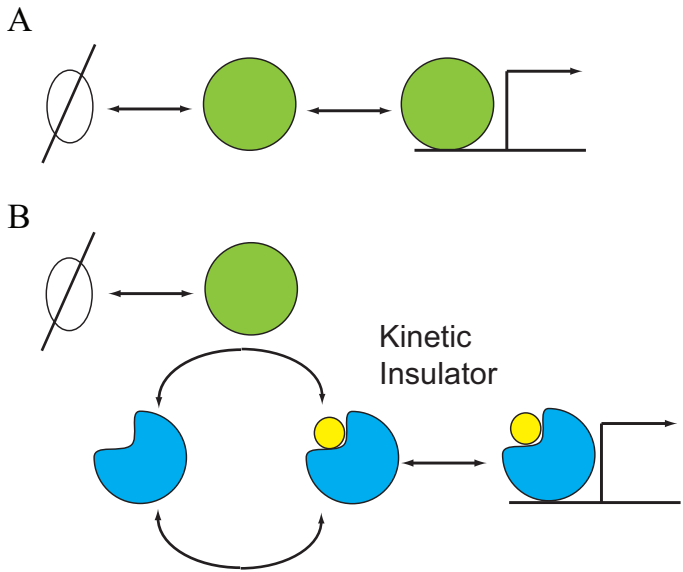}
\caption{{\bf Consuming energy to increase modularity.} (A) A transcription factor regulates downstream promoters. Sequestration of the transcription factor upon binding to promoters can lead to ``retroactivity'', i.e. a change in the dynamics of the transcription factor levels as a result of coupling to outputs. (B) Coupling the transcription factor through an insulating element consisting of a phosphorylation/dephosphorlyation cycle with fast dynamics reduces the effect of retroactivity.}
\label{Fig0}
\end{figure}

Another general role for energy consumption is to manipulate dynamics. By coupling a chemical reaction to energy sources such ATP or GTP, it is possible to change the dynamics of a biochemical network. One of the most interesting recent examples of how energy consumption can be used to change dynamics is the recent work of retroactivity\cite{Del-Vecchio2008Modular,Barton2013The-energy,Mishra2014A-load}. The central problem addressed in these papers is the observation that biochemical signal transduction circuits often have their dynamical behavior altered upon coupling to external outputs due to sequestration of proteins, a property dubbed ``retroactivity''. Such coupling is particularly undesired when there are a number of downstream outputs. These works demonstrate, both theoretically and experimentally, that it is possible to introduce insulating elements that reduce the magnitude of this retroactivity and thereby restore the modular dynamical behavior of synthetic circuits.  A key property of these insulating elements is that they utilize enzymatic futile cycles and hence actively consume energy. Moreover, a detailed theoretical analysis shows that the effectiveness of an insulating element is directly related to its energy consumption\cite{Barton2013The-energy}.
 
To demonstrate these concepts, we will consider the simple example of a protein $Z$ that is produced at a time-dependent rate $k(t)$ and is degraded at a rate $\delta$ (see Figure \ref{Fig0}). In addition, $Z$ regulates a family of promoters, with concentration $p_{\mathrm{tot}}$, by binding/unbinding to the promoter to form a complex $C$ at rates $k_{\mathrm{on/off}}$. The kinetics of this simple network is described by the set of ordinary differential equations
\bea
{dZ \over dt}&=& k(t) -\delta Z - \tau^{-1	}[k_{on}Z(p_{\mathrm{tot}}-C)+k_{\mathrm{off}} C], \nonumber\\
{dC \over dt} &=& \tau^{-1}[k_{\mathrm{on}} \tau Z(p_{\mathrm{tot}}-C)+k_{\mathrm{off}} \tau C],
\label{insulator}
\eea
where we have introduced an overall dimensionless timescale $\tau$ for the binding/unbinding dynamics. Notice that if $\tau^{-1} \gg 1$, then the timescale separation between the $Z$ and $C$ dynamics means that the $Z$ dynamics are well approximated by setting ${dC \over dt}=0$ so that 
\be
{dZ \over dt} \approx  k(t) -\delta Z.
\label{insulatorideal}
\ee
Thus, when $Z$ is coupled to a system with extremely fast dynamics, the retroactivity term, $\tau^{-1}[k_{on}Z(p_{\mathrm{tot}}-C)+k_{\mathrm{off}} C]$, is negligible.

This basic observation motivates the idea behind kinetic insulators. Instead of coupling $Z$ directly to the complex $C$, one couples $Z$ to $C$ indirectly through an intermediary insulating element with fast kinetics. Similar analysis of this more complex network shows that this dramatically decreases the amount of retroactivity. In practice, the insulating element is a phosphorylation/dephosphorylation cycle with fast kinetics (see Figure \ref{Fig0}). The faster the intermediary kinetics, and hence the more energy consumed by the futile cycle, the better the quasi-static approximation and the more effective the insulator (see \cite{Barton2013The-energy,Mishra2014A-load} for details).

\section{Reducing Variability}

Biochemical circuits can also consume energy to reduce variability and increase reproducibility. One of the best studied examples of this is the incredibly reproducible response of mammalian rod cells in response to light stimulation (see \cite{Bialek2012Biophysics-Searching} and references therein). This reproducibility of the rod cell response is especially surprising given that the response originates from the activation of a single rhodopsin molecule. A simple biophysically plausible model for an active rhodopsin is that its lifetime is exponentially distributed (i.e. the deactivation of rhodopsin is a Poisson process). In this case, the trial-to-trial variability, measured by the squared coefficient of variation, $CV^2=\sigma^2/\mu^2$, would be equal to $1$. Surprisingly, the actual  variability is much smaller than this naive expectation.

Experiments indicate that discrepancy is at least partially explained by the fact that the shut-off of active rhodopsin molecules proceeds through a multi-step cascade\cite{Rieke1996Molecular, Rieke1998Single-photon,Bialek2012Biophysics-Searching, Doan2006Multiple} (i.e the active rhodopsin molecule starts in state $1$, then transitions to state $2$, etc. until it reaches state $L$). If each of these steps were identical and independent, then from the central limit theorem the coefficient of variation of the $L$ step cascade would be $L$ times smaller than that of a single step, i.e.  $\sigma^2/\mu^2=1/L$. 

Notice that in order for such a multi-step cascade to reduce variability it is necessary that each of the transitions between the $L$ states be irreversible. If they were not, then one could not treat the $L$-steps as independent and the progression of the rhodopsin molecule through the various states would resemble a random walk, greatly increasing the variability \cite{Bialek2012Biophysics-Searching}. For this reason, reducing variability necessarily consumes energy. Consistent with this idea is the observation that the variability of rhodopsin seems to depend on the number of phosphorylation sites present on a rhodopsin molecule \cite{Doan2006Multiple}.

\begin{figure}[t]
\includegraphics[width=0.5\textwidth]{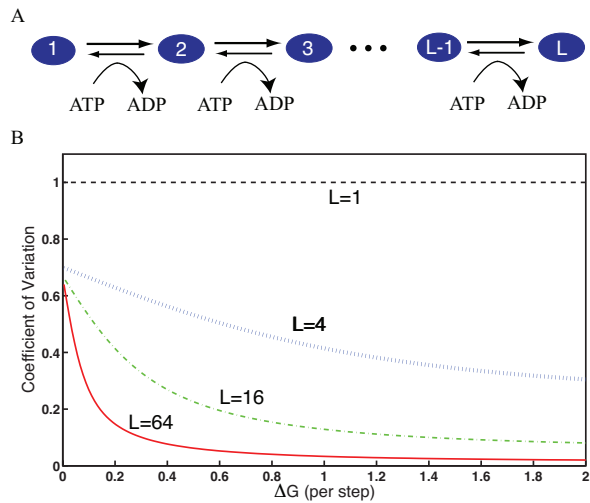}
\caption{{\bf Reducing variability in a multi-step cascade through energy consumption.} (A) A protein (blue ovals) is repeatedly phosphorylated $L$ times. (B) The coefficient of variation, defined as the variance over the mean squared of the time it takes to complete $L$ phosphorylations, as a function of the free-energy consumed during each step in the cascade, $\Delta G$, for $L=1,4,16,64$.}
\label{Fig1}
\end{figure}

In fact, it is possible to directly compute the coefficient of variation \cite{Munsky2009Specificity,Bel2010The-simplicity} as a function of the ratio of the forward and backward rates at each step, $\theta$. The logarithm of this ratio is simply the free-energy consumed at each step, $\Delta G= \log{\theta}$. Figure \ref{Fig1} shows that the coefficient of variation is a monotonically decreasing function of $\Delta G$ and hence the energy consumed by the cascade. Note that this decrease in the variability comes at the expense of a slower dynamic response, since the mean completion time scales linearly in the cascade length.  

Recent calculations have applied these ideas to the problem of a non-equilbrium receptor that estimates the concentration of an external ligand\cite{Lang2014Thermodynamics}. It was shown that by forcing the receptor to cycle through a series of $L$ states, one can increase the signal-to-noise ratio and construct a biochemical network that performs Maximum Likelihood Estimation (MLE) in the limit of large $L$. Since MLE is the statistically optimal estimator, this work suggest that it should be possible to improve the performance of synthetic biology based biodetectors by actively consuming energy.

Moreover, this trade-off between variability and energy consumption is likely to be quite general. Analytical arguments and numerical evidence suggest there may exist a general thermodynamic uncertainty relation relating the variance, of certain quantities in biochemical networks and the energy consumption \cite{Barato2015Thermodynamic}. In particular, achieving an uncertainty, $\sigma^2$, in a quantity such as the number of consumed/produced molecules in a genetic circuit or the number of steps in a molecular motor, requires an energetic cost of $2k_BT/\sigma^2$. This suggest that any strategy for reducing noise and variability in synthetic circuits will require these circuits to actively consume energy.

\section{Amplifying Signal}

Biochemical networks can also consume energy to amplify upstream input signals. Signal amplification is extremely important in many eukaryotic pathways designed to detect small changes in input such as the phototransduction pathway in the retina \cite{Detwiler2000Engineering} or the T cell receptor signaling pathway in immunology. In these pathways, a small change in the steady-state number of input messenger molecules, $dI$, leads to a large change in the steady-state number of output molecules, $dO$. The ratio of these changes is the number gain, often just called the gain,
\be
g_0 = {dO \over dI}
\ee
with $g_0>1$  implying the ratio of output to input molecules is necessarily greater than 1.

Before proceeding further, it is worth making the distinction between the number gain, which clearly measures changes in absolute number, with another commonly employed quantity used to describe biochemical pathways called logarithmic sensitivity \cite{Detwiler2000Engineering}. The logarithmic sensitivity, ${d\log{[O]} \over d \log{[I]}}$, measures the logarithmic change in the concentration of an output signal as a function of the logarithmic change in the input concentration and is a measure of the fractional or relative gain. Though logarithmic sensitivity and gain are often used interchangeably in the systems biology literature, the two measures are very different\cite{Detwiler2000Engineering}. To see this, consider a simple signaling element where a ligand, $L$ binds to a protein $X$ and changes its conformation to $X^*$. The input in this case is $L$ and the output is $X^*$. To have $g_0>1$, a small change in the number of ligands, $dL$ must produce a large change in the number of activated $X^*$. Notice that by definition, in equilibrium, ${dX^* \over dL} <1$ since each ligand can bind only one receptor. If instead $n$ ligands bind cooperatively to each $X$, then one would have ${dX^* \over dL} <1/n$. Thus, cooperatively in fact reduces the number gain. In contrast, the logarithmic sensitivity increases dramatically, ${d \log{[X]} \over d \log{[L]}}=n$.  An important consequence of this is that amplification of input signals (as measured by number gain) necessarily requires a non-equilibrium mechanism that consumes energy. 

The fact that energy consumption should be naturally related to the number gain and not logarithmic gain can be seen using both biological and physical arguments. The fundamental unit of energy is an ATP molecule. Since energy consumption is just a function of total number of ATP molecules hydrolyzed,  it is natural to measure gain using changes in the absolute numbers and not concentrations. From the viewpoints of physics, this is simply the statement that energy is an extensive quantity and hence depends on the actual number of molecules. 

\begin{figure}[t*]
\includegraphics[width=0.5\textwidth]{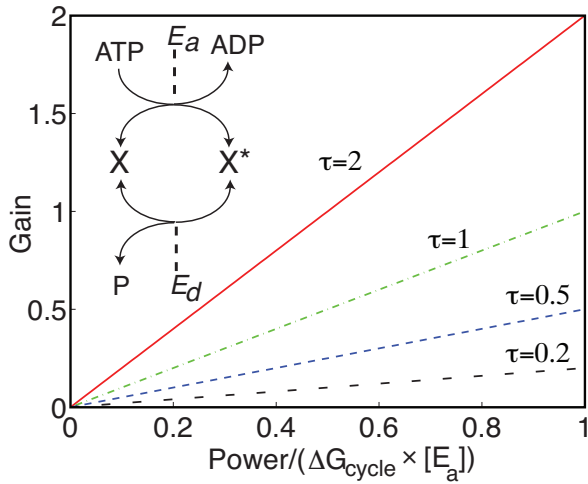}
\caption{{\bf Amplifying signals in a push-pull amplifier by consuming energy.} Schematic illustrates a simple push-pull amplifier where a kinase, $E_a$, modifies a protein from $X$ to $X^*$ and a phosphatase, $E_d$, catalyzing the reverse reaction. The plot illustrates that larger gain can be accomplished at the expense of a slower response time $\tau$.}
\label{Fig2}
\end{figure}

In biochemical networks, this signal amplification is accomplished through enzymatic cascades, where the input signal couples to an enzyme that can catalytically modify (e.g. phosphorylate) a substrate. Such basic enzymatic ``push-pull" amplifiers are the basic building block of many eukaryotic biochemical pathways, and are a canonical example of how energy consumption can be used to amplify input signals (see Figure \ref{Fig2}). A push-pull amplifier consists of an activating enzyme $E_a$ and a deactivating enzyme $E_d$ that interconvert a substrate between two forms, $X$ and $X^*$. Importantly, the post-translational modification of $X$ is coupled to a futile cycle such as ATP hydrolysis. The basic equations governing a push-pull amplifier are
\be
{d X^* \over dt} = \Gamma_a (E_a) X - \Gamma_d(E_d) X^*,
\label{kinetic}
\ee
where $\Gamma_a(E_a)$ is the rate at which enzyme $E_a$ converts $X$ to $X^*$ and $\Gamma_d(E_d)$ is the rate at which enzyme $E_d$ converts $X^*$ back to $X$. This rate equation must be supplemented by the conservation equation on the total number of $X$ molecules,
\be
X + X^* = X_{\mathrm{tot}}.
\ee

In the linear-response regime where the enzymes work far from saturation, one can approximate the rates in (\ref{kinetic}) as $ \Gamma_a (E_a) \approx k_a [E_a]$ and $\Gamma_d(E_d) \approx k_d [E_d]$, with $k_a= k_a^{\mathrm{cat}}/K_a$ and $k_d= k_d^{\mathrm{cat}}/K_d$ the ratios of the catalytic activity, $k_{cat}$, to the Michaelis-Menten constant, $K_M$, for the two enzymes. 
It is straightforward to show that the steady-state concentration of activated proteins is
\be
{\bar{X}^*} = \frac{X_{\mathrm{tot}} k_a [E_a]}{k_a[E_a]+k_d[E_d]}
\ee
Furthermore, one can define a ``response time'', $\tau$, for the enzymatic amplifier to be the rate at which a small perturbation from steady-state $\delta X^* = X^*-\bar{X^*}$ decays. This yields (see \cite{Detwiler2000Engineering} for details)
\be
\tau= (k_a[E_a]+k_d[E_d])^{-1}.
\ee
As discussed above, a key element of this enzymatic amplifier is that it works out of equilibrium. Each activation/deactivation event where the substrate cycles between the states  $X \mapsto X^* \mapsto X$ is coupled to a futile cycle (e.g. ATP hydrolysis) and hence dissipates an energy $\Delta G_{\mathrm cycle}$. At steady-state, the power consumption of the enzymatic amplifier is
\be
P=k_a [E_a] \bar{X}\Delta G_{\mathrm cycle}=k_d [E_d]\bar{X^*}\Delta G_{\mathrm cycle}.
\label{powerenzyme}
\ee

The input of the enzymatic amplifier is the number of activating enzymes $E_a$ and the output of the amplifier is the steady-state number of active substrate molecules $X^*$. This is natural in many eukaryotic signaling pathways where $E_a$ is often a receptor that becomes enzymatically active upon binding an external ligand. Using (\ref{powerenzyme}), one can calculate the static gain and find
\be
g_0=  (P/[E_a] ) \tau (\Delta G_{\mathrm cycle})^{-1}.
\ee
This expression shows that the gain of an enzymatic cascade is directly proportional to the power consumed per enzyme measured in the natural units of power that characterize the amplifier: $\Delta G_{\mathrm cycle}/\tau$. This is shown in Figure \ref{Fig2} where we plot the gain as a function of power consumption for different response times. 

Notice that the gain can be increased in two ways, by either increasing the power consumption or increasing the response time. Thus, at a fixed power consumption, increasing gain comes at the cost of a slower response. This is an example of a general engineering principle that is likely to be important for many applications in synthetic biology: the gain-bandwidth tradeoff \cite{Detwiler2000Engineering}. In general, a gain in signal comes at the expense of a reduced range of response frequencies (bandwidth). If one assumes that there is a maximum response frequency (ie a minimal time required for a response, a natural assumption in any practical engineering system), the gain-bandwidth tradeoff is equivalent to tradeoff between gain and response time. For this reason, energy consumption is likely to be an important consideration for synthetic circuits such as biosensors that must respond quickly to small changes in an external input. More generally, the gain-bandwidth tradeoff highlights the general tension between signal amplification, energy consumption, and signaling dynamics.

\section{Erasing Memory}
Memory is a central component of all computing devices. In a seminal 1961 paper\cite{Landauer1961Irreversibility}, Landauer outlined the fundamental thermodynamic and kinetic constraints that must be satisfied by memory modules in physical systems. Landauer emphasized the physical nature of information and used this to establish a connection between energy dissipation and erasing/resetting memory modules. This was codified in what is now known as LandauerÕs principle: any irreversible computing device must consume energy. 

The best understood example of a cellular computation from the perspective of statistical physics is the estimation of a steady-state concentration of chemical ligand in the surrounding environment by a biochemical network. This problem was first considered in the seminal paper\cite{Berg1977Physics} by Berg and Purcell who showed that the information a cell learns about its environment is limited by stochastic fluctuations in the occupancy of the receptors that detect the ligand. In particular, they considered the case of a cellular receptor that binds ligands at a concentration-dependent rate and unbinds particles at a fixed rate. They argued that cells could estimate chemical concentrations by calculating the average time a receptor is bound during a measurement time.

In these studies, the biochemical networks downstream of the receptors that perform the desired computations were largely ignored because the authors were interested in calculating fundamental limits on how well cells can estimate external concentrations. However, calculating energetic costs requires an explicit model of the downstream biochemical networks that implement these computations. As Feynman emphasized in his book on computation\cite{Feynman1998Feynman}, ``Information is physical.''

\begin{figure}[t!]
\includegraphics[width=0.5\textwidth]{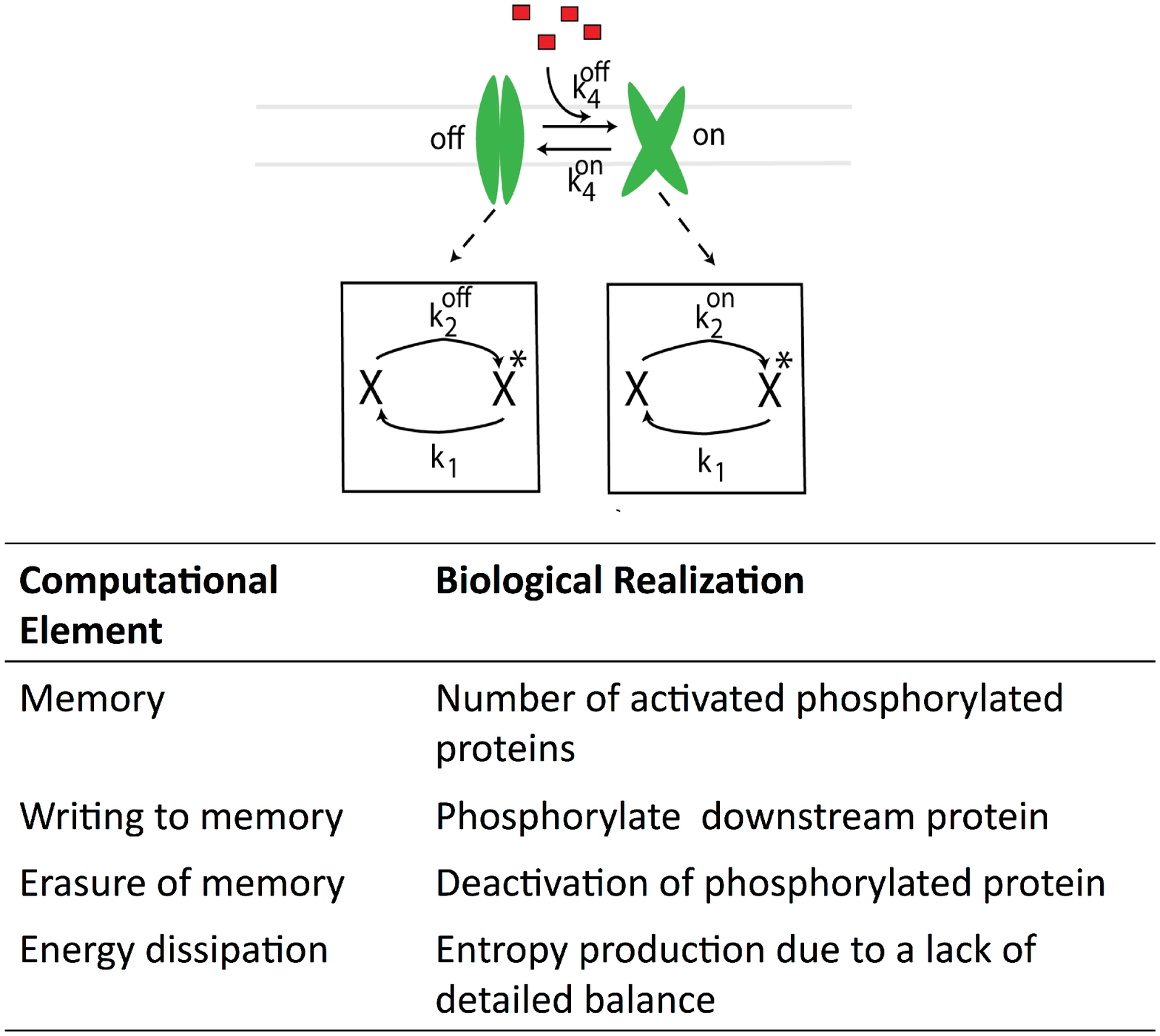}
\caption{{\bf A two-component network as a computational module.}  (A) Cellular network that calculates the Berg-Purcell statistic for estimating the concentration of an external-ligand. (B) Table summarizing the relationship between the network and standard computational elements and techniques.}
\label{Fig3}
\end{figure}

Recently, we considered a simple two-component biochemical network that directly computes the Berg-Purcell estimator\cite{Mehta2012Energetic}. Information about external ligand concentration is stored in the levels of a downstream protein (shown in Figure \ref{Fig3}). Such two-component networks are common in bacteria and are often used to sense external signals with receptors phosphorylating a downstream response regulator. Receptors convert a downstream protein from an inactive form to an active form at a state-dependent rate. The proteins are then inactivated at a state-independent rate. Interestingly, one can explicitly map components and functional operations in the network onto traditional computational tasks (see Figure \ref{Fig3}). Furthermore, it was shown that within the context of this network, computing the Berg-Purcell statistic necessarily required energy consumption. The underlying reason for this is that erasing/resetting memory requires energy (we note that while Landauer emphasized that erasing and not writing requires energy\cite{Landauer1961Irreversibility}, a recent paper argues energy consumption is bounded by writing to memory \cite{Sartori2014Thermodynamic}). These results seem to be quite general and similar conclusions have been reached by a variety of authors examining other biochemical networks \cite{Bo2015Thermodynamic,Govern2014Energy,Govern2014Optimal}.

These ideas have important implications for synthetic biology. Much as memory is central to the function of modern computers, biological memory modules are a crucial component of many synthetic gene circuits\cite{Bonnet2012Rewritable, Siuti2013Synthetic}. Any reusable synthetic circuit must possess a memory module that it can write and erase. Currently, synthetic circuits use two general classes of reusable memory modules that are commonly used: protein-based bistable genetic switches\cite{Gardner2000Construction} and recombinase-based DNA memory\cite{Bonnet2012Rewritable,Siuti2013Synthetic} (we are ignoring DNA mutation-based memories that can only be used once \cite{farzadfard2014genomically}). In both cases, resetting the memory involves consuming energy by expressing and degrading proteins (proteins involved in bistability and recombinases, respectively). Although this energy consumption is fundamental to any reusable memory module, it is desirable to find less energetically costly reusable memories that can still be stable over many generations such as chromatin-based memory module \cite{Keung2015Chromatin, keung2014using}. As synthetic circuits become increasingly complex, these energetic costs are likely to be ever more important.

\section{Using energy consumption to improve synthetic circuits}

Energy consumption is a defining feature of most information processing networks found in living systems. The theoretical work reviewed here provides new insights into biochemical networks. The greatest difference between equilibrium and non-equilibrium systems is that in equilibrium, the energy differences between states fundamentally determines the dynamics of the system, while in a non-equilibrium system the energy differences and dynamics become decoupled. This can be utilized in a variety of ways by biochemical networks and we broadly divided up the useful cases into relatively independent roles: increasing specificity, manipulating dynamics, reducing variability, amplifying signals, and erasing memory.  We believe that focusing on examples of each role will allow theorists and experimentalists to establish a common language and further both non-equilibrium physics and synthetic biology. One beautiful outcome of the interplay between theory and experiment is the recent work showing that a kinetic insulator that actively consumes energy can restore modularity and eliminate retroactivity in a simple synthetic circuit\cite{Mishra2014A-load}.

The theoretical results reviewed here can be summarized into several broad lessons on energy consumption that may prove useful for synthetic biology as well as providing theorists with future connections to experiments.
\begin{itemize}
\item \textbf{Fundamental Trade-Offs.} The ultimate limits of response speed, sensitivity, and energy consumption are in direct competition.
\item \textbf{Saturation of Trade-Offs.} Current works suggest that saturation effects are ubiquitous\cite{Murugan2012Speed,Murugan2014Discriminatory,Vaikuntanathan2014Dynamic} in energy consumption of biochemical networks and therefore only a few ATP may be enough\cite{Lang2014Thermodynamics} to nearly achieve the fundamental limits.
\item  \textbf{Futile Cycles are NOT Futile.} Futile cycles appear to be useless when only considering energy costs, but can provide benefits in terms of the fundamental trade-offs.
\item \textbf{Reusable Logic Must Consume Energy.} This is just the biological realization of Landauer's principle. Memory is especially important for circuits that function in stochastic environments where it is necessary to time-average over stochastic input signals.
\item \textbf{Chains are Useful.}  While it may seem redundant to have long chains of identical parts,  if the chain consumes energy this can improve specificity and reduce variation.
\item \textbf{Time Reversal Symmetry.} While equilibrium systems respect time reversal symmetry (forward and backwards flows are equivalent), energy consumption and non-equilibrium systems necessarily break this symmetry. This is especially important for synthetic circuit that seek to time-average stochastic inputs.
\item \textbf{Manipulate Time Scales.} Consuming energy can be useful to change the time scale of dynamics, as illustrated by the example of retroactivity and the introduction of energy consuming insulators.
\item \textbf{Information is Physical.} Theorists should heed Feynman's advice and attempt to translate theoretical advances into physical/biological devices.
\end{itemize}

We will end by focusing on one specific example that we believe is especially timely for synthetic biology. In naturally occurring biochemical networks, the primary source of energy for biochemical networks are futile cycles associated with post-translational modifications such as phosphorylation and methylation of residues. In contrast, energy dissipation in most synthetic circuits takes the form of the production and degradation of proteins. From the viewpoint of both energy and dynamics, protein degradation is an extremely inefficient solution to the problem. Proteins are metabolically expensive to synthesize, especially when compared to post-translational modifications. This may be one reason that most of the information processing and computation in eukaryotic signaling pathways is done through enzymatic cascades.

Designing synthetic circuits that can reap the full benefits of energy consumption requires developing new biological parts based around post-translational modifications of proteins. Such a ``post-transcriptional'' synthetic biology would allow to harness the manifold gains in performance that come from actively consuming energy without the extraordinary metabolic costs associated with protein synthesis. Currently, the power of this approach is limited by the dearth of circuit components that act at the level of post-translational modifications of proteins. Two promising directions that are seeking to overcome these limitations are phosphorylation-based synthetic signaling networks \cite{Lim2010Designing,Bashor2008Using, Wei2012Bacterial, Valk2014Multistep} and chromatin-based synthetic biology \cite{Keung2015Chromatin} that exploits reversible chromatin marks such as acetylation and methylation. In both cases, synthetic biologists are starting to engineer modular libraries of enzymes (kinases, phosphatases, chromatin reader-writers) to post-translationally modify specific protein substrates in response to particular signals. This will allow synthetic biology to take advantage of the increased information processing capabilities that arise from energy consumption.

\begin{acknowledgements}
 We would like to thank Caleb Bashor and Mo Khalil for many useful conversations and comments on the manuscript. PM was supported by a Simons Investigator in the Mathematical Modeling of Living Systems  grant and a Sloan Fellowship.  AHL was supported by a National Science Foundation Graduate Research Fellowship (NSF GRFP) under Grant No. DGE-1247312. DJS was supported by NIH Grant K25GM098875.
\end{acknowledgements} 


\begin{thebibliography}{82}%
\makeatletter
\providecommand \@ifxundefined [1]{%
 \@ifx{#1\undefined}
}%
\providecommand \@ifnum [1]{%
 \ifnum #1\expandafter \@firstoftwo
 \else \expandafter \@secondoftwo
 \fi
}%
\providecommand \@ifx [1]{%
 \ifx #1\expandafter \@firstoftwo
 \else \expandafter \@secondoftwo
 \fi
}%
\providecommand \natexlab [1]{#1}%
\providecommand \enquote  [1]{``#1''}%
\providecommand \bibnamefont  [1]{#1}%
\providecommand \bibfnamefont [1]{#1}%
\providecommand \citenamefont [1]{#1}%
\providecommand \href@noop [0]{\@secondoftwo}%
\providecommand \href [0]{\begingroup \@sanitize@url \@href}%
\providecommand \@href[1]{\@@startlink{#1}\@@href}%
\providecommand \@@href[1]{\endgroup#1\@@endlink}%
\providecommand \@sanitize@url [0]{\catcode `\\12\catcode `\$12\catcode
  `\&12\catcode `\#12\catcode `\^12\catcode `\_12\catcode `\%12\relax}%
\providecommand \@@startlink[1]{}%
\providecommand \@@endlink[0]{}%
\providecommand \url  [0]{\begingroup\@sanitize@url \@url }%
\providecommand \@url [1]{\endgroup\@href {#1}{\urlprefix }}%
\providecommand \urlprefix  [0]{URL }%
\providecommand \Eprint [0]{\href }%
\providecommand \doibase [0]{http://dx.doi.org/}%
\providecommand \selectlanguage [0]{\@gobble}%
\providecommand \bibinfo  [0]{\@secondoftwo}%
\providecommand \bibfield  [0]{\@secondoftwo}%
\providecommand \translation [1]{[#1]}%
\providecommand \BibitemOpen [0]{}%
\providecommand \bibitemStop [0]{}%
\providecommand \bibitemNoStop [0]{.\EOS\space}%
\providecommand \EOS [0]{\spacefactor3000\relax}%
\providecommand \BibitemShut  [1]{\csname bibitem#1\endcsname}%
\let\auto@bib@innerbib\@empty
\bibitem [{\citenamefont {Miller}\ and\ \citenamefont
  {Bassler}(2001)}]{Miller2001Quorum}%
  \BibitemOpen
  \bibfield  {author} {\bibinfo {author} {\bibfnamefont {M.~B.}\ \bibnamefont
  {Miller}}\ and\ \bibinfo {author} {\bibfnamefont {B.~L.}\ \bibnamefont
  {Bassler}},\ }\href {\doibase 10.1146/annurev.micro.55.1.165} {\bibfield
  {journal} {\bibinfo  {journal} {Annual Review of Microbiology}\ }\textbf
  {\bibinfo {volume} {55}},\ \bibinfo {pages} {165} (\bibinfo {year}
  {2001})}\BibitemShut {NoStop}%
\bibitem [{\citenamefont {Sourjik}\ and\ \citenamefont
  {Wingreen}(2012)}]{Sourjik2012Responding}%
  \BibitemOpen
  \bibfield  {author} {\bibinfo {author} {\bibfnamefont {V.}~\bibnamefont
  {Sourjik}}\ and\ \bibinfo {author} {\bibfnamefont {N.~S.}\ \bibnamefont
  {Wingreen}},\ }\href {\doibase http://dx.doi.org/10.1016/j.ceb.2011.11.008}
  {\bibfield  {journal} {\bibinfo  {journal} {Current Opinion in Cell Biology}\
  }\textbf {\bibinfo {volume} {24}},\ \bibinfo {pages} {262} (\bibinfo {year}
  {2012})}\BibitemShut {NoStop}%
\bibitem [{\citenamefont {Logan}\ and\ \citenamefont
  {Nusse}(2004)}]{Logan2004The-WNT-Signaling}%
  \BibitemOpen
  \bibfield  {author} {\bibinfo {author} {\bibfnamefont {C.~Y.}\ \bibnamefont
  {Logan}}\ and\ \bibinfo {author} {\bibfnamefont {R.}~\bibnamefont {Nusse}},\
  }\href {\doibase 10.1146/annurev.cellbio.20.010403.113126} {\bibfield
  {journal} {\bibinfo  {journal} {Annual Review of Cell and Developmental
  Biology}\ }\textbf {\bibinfo {volume} {20}},\ \bibinfo {pages} {781}
  (\bibinfo {year} {2004})}\BibitemShut {NoStop}%
\bibitem [{\citenamefont {Gardner}\ \emph {et~al.}(2000)\citenamefont
  {Gardner}, \citenamefont {Cantor},\ and\ \citenamefont
  {Collins}}]{Gardner2000Construction}%
  \BibitemOpen
  \bibfield  {author} {\bibinfo {author} {\bibfnamefont {T.~S.}\ \bibnamefont
  {Gardner}}, \bibinfo {author} {\bibfnamefont {C.~R.}\ \bibnamefont {Cantor}},
  \ and\ \bibinfo {author} {\bibfnamefont {J.~J.}\ \bibnamefont {Collins}},\
  }\href {http://dx.doi.org/10.1038/35002131} {\bibfield  {journal} {\bibinfo
  {journal} {Nature}\ }\textbf {\bibinfo {volume} {403}},\ \bibinfo {pages}
  {339} (\bibinfo {year} {2000})}\BibitemShut {NoStop}%
\bibitem [{\citenamefont {Elowitz}\ and\ \citenamefont
  {Leibler}(2000)}]{Elowitz2000A-synthetic}%
  \BibitemOpen
  \bibfield  {author} {\bibinfo {author} {\bibfnamefont {M.~B.}\ \bibnamefont
  {Elowitz}}\ and\ \bibinfo {author} {\bibfnamefont {S.}~\bibnamefont
  {Leibler}},\ }\href@noop {} {\bibfield  {journal} {\bibinfo  {journal}
  {Nature}\ }\textbf {\bibinfo {volume} {403}},\ \bibinfo {pages} {335}
  (\bibinfo {year} {2000})}\BibitemShut {NoStop}%
\bibitem [{\citenamefont {Friedland}\ \emph {et~al.}(2009)\citenamefont
  {Friedland}, \citenamefont {Lu}, \citenamefont {Wang}, \citenamefont {Shi},
  \citenamefont {Church},\ and\ \citenamefont
  {Collins}}]{Friedland2009Synthetic}%
  \BibitemOpen
  \bibfield  {author} {\bibinfo {author} {\bibfnamefont {A.~E.}\ \bibnamefont
  {Friedland}}, \bibinfo {author} {\bibfnamefont {T.~K.}\ \bibnamefont {Lu}},
  \bibinfo {author} {\bibfnamefont {X.}~\bibnamefont {Wang}}, \bibinfo {author}
  {\bibfnamefont {D.}~\bibnamefont {Shi}}, \bibinfo {author} {\bibfnamefont
  {G.}~\bibnamefont {Church}}, \ and\ \bibinfo {author} {\bibfnamefont {J.~J.}\
  \bibnamefont {Collins}},\ }\href
  {http://www.sciencemag.org/content/324/5931/1199.abstract} {\bibfield
  {journal} {\bibinfo  {journal} {Science}\ }\textbf {\bibinfo {volume}
  {324}},\ \bibinfo {pages} {1199} (\bibinfo {year} {2009})}\BibitemShut
  {NoStop}%
\bibitem [{\citenamefont {Danino}\ \emph {et~al.}(2010)\citenamefont {Danino},
  \citenamefont {Mondragon-Palomino}, \citenamefont {Tsimring},\ and\
  \citenamefont {Hasty}}]{Danino2010A-synchronized}%
  \BibitemOpen
  \bibfield  {author} {\bibinfo {author} {\bibfnamefont {T.}~\bibnamefont
  {Danino}}, \bibinfo {author} {\bibfnamefont {O.}~\bibnamefont
  {Mondragon-Palomino}}, \bibinfo {author} {\bibfnamefont {L.}~\bibnamefont
  {Tsimring}}, \ and\ \bibinfo {author} {\bibfnamefont {J.}~\bibnamefont
  {Hasty}},\ }\href {http://dx.doi.org/10.1038/nature08753} {\bibfield
  {journal} {\bibinfo  {journal} {Nature}\ }\textbf {\bibinfo {volume} {463}},\
  \bibinfo {pages} {326} (\bibinfo {year} {2010})}\BibitemShut {NoStop}%
\bibitem [{\citenamefont {Mukherji}\ and\ \citenamefont {van
  Oudenaarden}(2009)}]{Mukherji2009Synthetic}%
  \BibitemOpen
  \bibfield  {author} {\bibinfo {author} {\bibfnamefont {S.}~\bibnamefont
  {Mukherji}}\ and\ \bibinfo {author} {\bibfnamefont {A.}~\bibnamefont {van
  Oudenaarden}},\ }\href {http://dx.doi.org/10.1038/nrg2697} {\bibfield
  {journal} {\bibinfo  {journal} {Nat Rev Genet}\ }\textbf {\bibinfo {volume}
  {10}},\ \bibinfo {pages} {859} (\bibinfo {year} {2009})}\BibitemShut
  {NoStop}%
\bibitem [{\citenamefont {Purnick}\ and\ \citenamefont
  {Weiss}(2009)}]{Purnick2009The-second}%
  \BibitemOpen
  \bibfield  {author} {\bibinfo {author} {\bibfnamefont {P.~E.~M.}\
  \bibnamefont {Purnick}}\ and\ \bibinfo {author} {\bibfnamefont
  {R.}~\bibnamefont {Weiss}},\ }\href {http://dx.doi.org/10.1038/nrm2698}
  {\bibfield  {journal} {\bibinfo  {journal} {Nat Rev Mol Cell Biol}\ }\textbf
  {\bibinfo {volume} {10}},\ \bibinfo {pages} {410} (\bibinfo {year}
  {2009})}\BibitemShut {NoStop}%
\bibitem [{\citenamefont {Khalil}\ and\ \citenamefont
  {Collins}(2010)}]{Khalil2010Synthetic}%
  \BibitemOpen
  \bibfield  {author} {\bibinfo {author} {\bibfnamefont {A.~S.}\ \bibnamefont
  {Khalil}}\ and\ \bibinfo {author} {\bibfnamefont {J.~J.}\ \bibnamefont
  {Collins}},\ }\href {http://dx.doi.org/10.1038/nrg2775} {\bibfield  {journal}
  {\bibinfo  {journal} {Nat Rev Genet}\ }\textbf {\bibinfo {volume} {11}},\
  \bibinfo {pages} {367} (\bibinfo {year} {2010})}\BibitemShut {NoStop}%
\bibitem [{\citenamefont {Nandagopal}\ and\ \citenamefont
  {Elowitz}(2011)}]{Nandagopal2011Synthetic}%
  \BibitemOpen
  \bibfield  {author} {\bibinfo {author} {\bibfnamefont {N.}~\bibnamefont
  {Nandagopal}}\ and\ \bibinfo {author} {\bibfnamefont {M.~B.}\ \bibnamefont
  {Elowitz}},\ }\href
  {http://www.sciencemag.org/content/333/6047/1244.abstract} {\bibfield
  {journal} {\bibinfo  {journal} {Science}\ }\textbf {\bibinfo {volume}
  {333}},\ \bibinfo {pages} {1244} (\bibinfo {year} {2011})}\BibitemShut
  {NoStop}%
\bibitem [{\citenamefont {Ruder}\ \emph {et~al.}(2011)\citenamefont {Ruder},
  \citenamefont {Lu},\ and\ \citenamefont {Collins}}]{Ruder2011Synthetic}%
  \BibitemOpen
  \bibfield  {author} {\bibinfo {author} {\bibfnamefont {W.~C.}\ \bibnamefont
  {Ruder}}, \bibinfo {author} {\bibfnamefont {T.}~\bibnamefont {Lu}}, \ and\
  \bibinfo {author} {\bibfnamefont {J.~J.}\ \bibnamefont {Collins}},\ }\href
  {http://www.sciencemag.org/content/333/6047/1248.abstract} {\bibfield
  {journal} {\bibinfo  {journal} {Science}\ }\textbf {\bibinfo {volume}
  {333}},\ \bibinfo {pages} {1248} (\bibinfo {year} {2011})}\BibitemShut
  {NoStop}%
\bibitem [{\citenamefont {Keung}\ \emph {et~al.}(2015)\citenamefont {Keung},
  \citenamefont {Joung}, \citenamefont {Khalil},\ and\ \citenamefont
  {Collins}}]{Keung2015Chromatin}%
  \BibitemOpen
  \bibfield  {author} {\bibinfo {author} {\bibfnamefont {A.~J.}\ \bibnamefont
  {Keung}}, \bibinfo {author} {\bibfnamefont {J.~K.}\ \bibnamefont {Joung}},
  \bibinfo {author} {\bibfnamefont {A.~S.}\ \bibnamefont {Khalil}}, \ and\
  \bibinfo {author} {\bibfnamefont {J.~J.}\ \bibnamefont {Collins}},\ }\href
  {http://dx.doi.org/10.1038/nrg3900} {\bibfield  {journal} {\bibinfo
  {journal} {Nat Rev Genet}\ }\textbf {\bibinfo {volume} {16}},\ \bibinfo
  {pages} {159} (\bibinfo {year} {2015})}\BibitemShut {NoStop}%
\bibitem [{\citenamefont {Cameron}\ \emph {et~al.}(2014)\citenamefont
  {Cameron}, \citenamefont {Bashor},\ and\ \citenamefont
  {Collins}}]{Cameron2014A-brief}%
  \BibitemOpen
  \bibfield  {author} {\bibinfo {author} {\bibfnamefont {D.~E.}\ \bibnamefont
  {Cameron}}, \bibinfo {author} {\bibfnamefont {C.~J.}\ \bibnamefont {Bashor}},
  \ and\ \bibinfo {author} {\bibfnamefont {J.~J.}\ \bibnamefont {Collins}},\
  }\href {http://dx.doi.org/10.1038/nrmicro3239} {\bibfield  {journal}
  {\bibinfo  {journal} {Nat Rev Micro}\ }\textbf {\bibinfo {volume} {12}},\
  \bibinfo {pages} {381} (\bibinfo {year} {2014})}\BibitemShut {NoStop}%
\bibitem [{\citenamefont {Cheng}\ and\ \citenamefont
  {Lu}(2012)}]{Cheng2012Synthetic}%
  \BibitemOpen
  \bibfield  {author} {\bibinfo {author} {\bibfnamefont {A.~A.}\ \bibnamefont
  {Cheng}}\ and\ \bibinfo {author} {\bibfnamefont {T.~K.}\ \bibnamefont {Lu}},\
  }\href {\doibase 10.1146/annurev-bioeng-071811-150118} {\bibfield  {journal}
  {\bibinfo  {journal} {Annual Review of Biomedical Engineering}\ }\textbf
  {\bibinfo {volume} {14}},\ \bibinfo {pages} {155} (\bibinfo {year}
  {2012})}\BibitemShut {NoStop}%
\bibitem [{\citenamefont {Way}\ \emph {et~al.}(2014)\citenamefont {Way},
  \citenamefont {Collins}, \citenamefont {Keasling},\ and\ \citenamefont
  {Silver}}]{Way2014Integrating}%
  \BibitemOpen
  \bibfield  {author} {\bibinfo {author} {\bibfnamefont {J.~C.}\ \bibnamefont
  {Way}}, \bibinfo {author} {\bibfnamefont {J.~J.}\ \bibnamefont {Collins}},
  \bibinfo {author} {\bibfnamefont {J.~D.}\ \bibnamefont {Keasling}}, \ and\
  \bibinfo {author} {\bibfnamefont {P.~A.}\ \bibnamefont {Silver}},\
  }\href@noop {} {\bibfield  {journal} {\bibinfo  {journal} {Cell}\ }\textbf
  {\bibinfo {volume} {157}},\ \bibinfo {pages} {151} (\bibinfo {year}
  {2014})}\BibitemShut {NoStop}%
\bibitem [{\citenamefont {Kwok}(2010)}]{Kwok2010Five}%
  \BibitemOpen
  \bibfield  {author} {\bibinfo {author} {\bibfnamefont {R.}~\bibnamefont
  {Kwok}},\ }\href@noop {} {\bibfield  {journal} {\bibinfo  {journal} {Nature
  News}\ }\textbf {\bibinfo {volume} {463}},\ \bibinfo {pages} {288} (\bibinfo
  {year} {2010})}\BibitemShut {NoStop}%
\bibitem [{\citenamefont {Landauer}(1961)}]{Landauer1961Irreversibility}%
  \BibitemOpen
  \bibfield  {author} {\bibinfo {author} {\bibfnamefont {R.}~\bibnamefont
  {Landauer}},\ }\href@noop {} {\bibfield  {journal} {\bibinfo  {journal} {IBM
  Journal of Research and Development}\ }\textbf {\bibinfo {volume} {5}},\
  \bibinfo {pages} {183} (\bibinfo {year} {1961})}\BibitemShut {NoStop}%
\bibitem [{\citenamefont {Laughlin}(2001)}]{Laughlin2001Energy}%
  \BibitemOpen
  \bibfield  {author} {\bibinfo {author} {\bibfnamefont {S.~B.}\ \bibnamefont
  {Laughlin}},\ }\href@noop {} {\bibfield  {journal} {\bibinfo  {journal}
  {Current opinion in neurobiology}\ }\textbf {\bibinfo {volume} {11}},\
  \bibinfo {pages} {475} (\bibinfo {year} {2001})}\BibitemShut {NoStop}%
\bibitem [{\citenamefont {Qian}\ and\ \citenamefont
  {Beard}(2005)}]{Qian2005Thermodynamics}%
  \BibitemOpen
  \bibfield  {author} {\bibinfo {author} {\bibfnamefont {H.}~\bibnamefont
  {Qian}}\ and\ \bibinfo {author} {\bibfnamefont {D.~A.}\ \bibnamefont
  {Beard}},\ }\href {\doibase http://dx.doi.org/10.1016/j.bpc.2004.12.001}
  {\bibfield  {journal} {\bibinfo  {journal} {Biophysical Chemistry}\ }\textbf
  {\bibinfo {volume} {114}},\ \bibinfo {pages} {213} (\bibinfo {year}
  {2005})}\BibitemShut {NoStop}%
\bibitem [{\citenamefont {Qian}(2007)}]{Qian2007Phosphorylation}%
  \BibitemOpen
  \bibfield  {author} {\bibinfo {author} {\bibfnamefont {H.}~\bibnamefont
  {Qian}},\ }\href
  {http://www.annualreviews.org/doi/abs/10.1146/annurev.physchem.58.032806.104550}
  {\bibfield  {journal} {\bibinfo  {journal} {Annual Review of Physical
  Chemistry}\ }\textbf {\bibinfo {volume} {58}},\ \bibinfo {pages} {113}
  (\bibinfo {year} {2007})}\BibitemShut {NoStop}%
\bibitem [{\citenamefont {Zhang}\ \emph {et~al.}(2012)\citenamefont {Zhang},
  \citenamefont {Qian},\ and\ \citenamefont {Qian}}]{Zhang2012Stochastic}%
  \BibitemOpen
  \bibfield  {author} {\bibinfo {author} {\bibfnamefont {X.-J.}\ \bibnamefont
  {Zhang}}, \bibinfo {author} {\bibfnamefont {H.}~\bibnamefont {Qian}}, \ and\
  \bibinfo {author} {\bibfnamefont {M.}~\bibnamefont {Qian}},\ }\href {\doibase
  http://dx.doi.org/10.1016/j.physrep.2011.09.002} {\bibfield  {journal}
  {\bibinfo  {journal} {Physics Reports}\ }\textbf {\bibinfo {volume} {510}},\
  \bibinfo {pages} {1} (\bibinfo {year} {2012})}\BibitemShut {NoStop}%
\bibitem [{\citenamefont {Ge}\ \emph {et~al.}(2012)\citenamefont {Ge},
  \citenamefont {Qian},\ and\ \citenamefont {Qian}}]{Ge2012Stochastic}%
  \BibitemOpen
  \bibfield  {author} {\bibinfo {author} {\bibfnamefont {H.}~\bibnamefont
  {Ge}}, \bibinfo {author} {\bibfnamefont {M.}~\bibnamefont {Qian}}, \ and\
  \bibinfo {author} {\bibfnamefont {H.}~\bibnamefont {Qian}},\ }\href {\doibase
  http://dx.doi.org/10.1016/j.physrep.2011.09.001} {\bibfield  {journal}
  {\bibinfo  {journal} {Physics Reports}\ }\textbf {\bibinfo {volume} {510}},\
  \bibinfo {pages} {87} (\bibinfo {year} {2012})}\BibitemShut {NoStop}%
\bibitem [{\citenamefont {Barato}\ \emph
  {et~al.}(2013{\natexlab{a}})\citenamefont {Barato}, \citenamefont {Hartich},\
  and\ \citenamefont {Seifert}}]{Barato2013Information-theoretic}%
  \BibitemOpen
  \bibfield  {author} {\bibinfo {author} {\bibfnamefont {A.}~\bibnamefont
  {Barato}}, \bibinfo {author} {\bibfnamefont {D.}~\bibnamefont {Hartich}}, \
  and\ \bibinfo {author} {\bibfnamefont {U.}~\bibnamefont {Seifert}},\
  }\href@noop {} {\bibfield  {journal} {\bibinfo  {journal} {Physical Review
  E}\ }\textbf {\bibinfo {volume} {87}},\ \bibinfo {pages} {042104} (\bibinfo
  {year} {2013}{\natexlab{a}})}\BibitemShut {NoStop}%
\bibitem [{\citenamefont {Lan}\ \emph {et~al.}(2012)\citenamefont {Lan},
  \citenamefont {Sartori}, \citenamefont {Neumann}, \citenamefont {Sourjik},\
  and\ \citenamefont {Tu}}]{Lan2012The-energy-speed-accuracy}%
  \BibitemOpen
  \bibfield  {author} {\bibinfo {author} {\bibfnamefont {G.}~\bibnamefont
  {Lan}}, \bibinfo {author} {\bibfnamefont {P.}~\bibnamefont {Sartori}},
  \bibinfo {author} {\bibfnamefont {S.}~\bibnamefont {Neumann}}, \bibinfo
  {author} {\bibfnamefont {V.}~\bibnamefont {Sourjik}}, \ and\ \bibinfo
  {author} {\bibfnamefont {Y.}~\bibnamefont {Tu}},\ }\href
  {http://dx.doi.org/10.1038/nphys2276} {\bibfield  {journal} {\bibinfo
  {journal} {Nat Phys}\ }\textbf {\bibinfo {volume} {8}},\ \bibinfo {pages}
  {422} (\bibinfo {year} {2012})}\BibitemShut {NoStop}%
\bibitem [{\citenamefont {Govern}\ and\ \citenamefont {ten
  Wolde}(2012)}]{Govern2012Fundamental}%
  \BibitemOpen
  \bibfield  {author} {\bibinfo {author} {\bibfnamefont {C.~C.}\ \bibnamefont
  {Govern}}\ and\ \bibinfo {author} {\bibfnamefont {P.~R.}\ \bibnamefont {ten
  Wolde}},\ }\href {http://link.aps.org/doi/10.1103/PhysRevLett.109.218103}
  {\bibfield  {journal} {\bibinfo  {journal} {Physical Review Letters}\
  }\textbf {\bibinfo {volume} {109}},\ \bibinfo {pages} {218103} (\bibinfo
  {year} {2012})}\BibitemShut {NoStop}%
\bibitem [{\citenamefont {Sagawa}\ and\ \citenamefont
  {Ueda}(2012)}]{Sagawa2012Nonequilibrium}%
  \BibitemOpen
  \bibfield  {author} {\bibinfo {author} {\bibfnamefont {T.}~\bibnamefont
  {Sagawa}}\ and\ \bibinfo {author} {\bibfnamefont {M.}~\bibnamefont {Ueda}},\
  }\href {http://link.aps.org/doi/10.1103/PhysRevE.85.021104} {\bibfield
  {journal} {\bibinfo  {journal} {Physical Review E}\ }\textbf {\bibinfo
  {volume} {85}},\ \bibinfo {pages} {021104} (\bibinfo {year}
  {2012})}\BibitemShut {NoStop}%
\bibitem [{\citenamefont {Still}\ \emph {et~al.}(2012)\citenamefont {Still},
  \citenamefont {Sivak}, \citenamefont {Bell},\ and\ \citenamefont
  {Crooks}}]{Still2012Thermodynamics}%
  \BibitemOpen
  \bibfield  {author} {\bibinfo {author} {\bibfnamefont {S.}~\bibnamefont
  {Still}}, \bibinfo {author} {\bibfnamefont {D.~A.}\ \bibnamefont {Sivak}},
  \bibinfo {author} {\bibfnamefont {A.~J.}\ \bibnamefont {Bell}}, \ and\
  \bibinfo {author} {\bibfnamefont {G.~E.}\ \bibnamefont {Crooks}},\ }\href
  {http://link.aps.org/doi/10.1103/PhysRevLett.109.120604} {\bibfield
  {journal} {\bibinfo  {journal} {Physical Review Letters}\ }\textbf {\bibinfo
  {volume} {109}},\ \bibinfo {pages} {120604} (\bibinfo {year}
  {2012})}\BibitemShut {NoStop}%
\bibitem [{\citenamefont {Barato}\ \emph
  {et~al.}(2013{\natexlab{b}})\citenamefont {Barato}, \citenamefont {Hartich},\
  and\ \citenamefont {Seifert}}]{Barato2013Information}%
  \BibitemOpen
  \bibfield  {author} {\bibinfo {author} {\bibfnamefont {A.~C.}\ \bibnamefont
  {Barato}}, \bibinfo {author} {\bibfnamefont {D.}~\bibnamefont {Hartich}}, \
  and\ \bibinfo {author} {\bibfnamefont {U.}~\bibnamefont {Seifert}},\ }\href
  {http://link.aps.org/doi/10.1103/PhysRevE.87.042104} {\bibfield  {journal}
  {\bibinfo  {journal} {Physical Review E}\ }\textbf {\bibinfo {volume} {87}},\
  \bibinfo {pages} {042104} (\bibinfo {year} {2013}{\natexlab{b}})}\BibitemShut
  {NoStop}%
\bibitem [{\citenamefont {Govern}\ and\ \citenamefont
  {Wolde}(2013)}]{Govern2013How-biochemical}%
  \BibitemOpen
  \bibfield  {author} {\bibinfo {author} {\bibfnamefont {C.~C.}\ \bibnamefont
  {Govern}}\ and\ \bibinfo {author} {\bibfnamefont {P.~R.~t.}\ \bibnamefont
  {Wolde}},\ }\href@noop {} {\bibfield  {journal} {\bibinfo  {journal} {arXiv
  preprint arXiv:1308.1449}\ } (\bibinfo {year} {2013})}\BibitemShut {NoStop}%
\bibitem [{\citenamefont {Becker}\ \emph {et~al.}(2013)\citenamefont {Becker},
  \citenamefont {Mugler},\ and\ \citenamefont {Wolde}}]{Becker2013Prediction}%
  \BibitemOpen
  \bibfield  {author} {\bibinfo {author} {\bibfnamefont {N.~B.}\ \bibnamefont
  {Becker}}, \bibinfo {author} {\bibfnamefont {A.}~\bibnamefont {Mugler}}, \
  and\ \bibinfo {author} {\bibfnamefont {P.~R.~t.}\ \bibnamefont {Wolde}},\
  }\href@noop {} {\bibfield  {journal} {\bibinfo  {journal} {arXiv preprint
  arXiv:1312.5625}\ } (\bibinfo {year} {2013})}\BibitemShut {NoStop}%
\bibitem [{\citenamefont {Bo}\ \emph {et~al.}(2015)\citenamefont {Bo},
  \citenamefont {Giudice},\ and\ \citenamefont {Celani}}]{Bo2015Thermodynamic}%
  \BibitemOpen
  \bibfield  {author} {\bibinfo {author} {\bibfnamefont {S.}~\bibnamefont
  {Bo}}, \bibinfo {author} {\bibfnamefont {M.~D.}\ \bibnamefont {Giudice}}, \
  and\ \bibinfo {author} {\bibfnamefont {A.}~\bibnamefont {Celani}},\ }\href
  {http://stacks.iop.org/1742-5468/2015/i=1/a=P01014} {\bibfield  {journal}
  {\bibinfo  {journal} {Journal of Statistical Mechanics: Theory and
  Experiment}\ }\textbf {\bibinfo {volume} {2015}},\ \bibinfo {pages} {P01014}
  (\bibinfo {year} {2015})}\BibitemShut {NoStop}%
\bibitem [{\citenamefont {Govern}\ and\ \citenamefont {ten
  Wolde}(2014{\natexlab{a}})}]{Govern2014Energy}%
  \BibitemOpen
  \bibfield  {author} {\bibinfo {author} {\bibfnamefont {C.~C.}\ \bibnamefont
  {Govern}}\ and\ \bibinfo {author} {\bibfnamefont {P.~R.}\ \bibnamefont {ten
  Wolde}},\ }\href {http://link.aps.org/doi/10.1103/PhysRevLett.113.258102}
  {\bibfield  {journal} {\bibinfo  {journal} {Physical Review Letters}\
  }\textbf {\bibinfo {volume} {113}},\ \bibinfo {pages} {258102} (\bibinfo
  {year} {2014}{\natexlab{a}})}\BibitemShut {NoStop}%
\bibitem [{\citenamefont {Govern}\ and\ \citenamefont {ten
  Wolde}(2014{\natexlab{b}})}]{Govern2014Optimal}%
  \BibitemOpen
  \bibfield  {author} {\bibinfo {author} {\bibfnamefont {C.~C.}\ \bibnamefont
  {Govern}}\ and\ \bibinfo {author} {\bibfnamefont {P.~R.}\ \bibnamefont {ten
  Wolde}},\ }\href
  {http://www.pnas.org/content/early/2014/11/21/1411524111.abstract} {\bibfield
   {journal} {\bibinfo  {journal} {Proceedings of the National Academy of
  Sciences}\ } (\bibinfo {year} {2014}{\natexlab{b}})}\BibitemShut {NoStop}%
\bibitem [{\citenamefont {Kaizu}\ \emph {et~al.}(2014)\citenamefont {Kaizu},
  \citenamefont {de~Ronde}, \citenamefont {Paijmans}, \citenamefont
  {Takahashi}, \citenamefont {Tostevin},\ and\ \citenamefont {ten
  Wolde}}]{Kaizu2014Berg}%
  \BibitemOpen
  \bibfield  {author} {\bibinfo {author} {\bibfnamefont {K.}~\bibnamefont
  {Kaizu}}, \bibinfo {author} {\bibfnamefont {W.}~\bibnamefont {de~Ronde}},
  \bibinfo {author} {\bibfnamefont {J.}~\bibnamefont {Paijmans}}, \bibinfo
  {author} {\bibfnamefont {K.}~\bibnamefont {Takahashi}}, \bibinfo {author}
  {\bibfnamefont {F.}~\bibnamefont {Tostevin}}, \ and\ \bibinfo {author}
  {\bibfnamefont {P.~R.}\ \bibnamefont {ten Wolde}},\ }\href {\doibase
  http://dx.doi.org/10.1016/j.bpj.2013.12.030} {\bibfield  {journal} {\bibinfo
  {journal} {Biophysical Journal}\ }\textbf {\bibinfo {volume} {106}},\
  \bibinfo {pages} {976} (\bibinfo {year} {2014})}\BibitemShut {NoStop}%
\bibitem [{\citenamefont {Iyengar}\ and\ \citenamefont
  {Rao}(2014)}]{Iyengar2014A-cellular}%
  \BibitemOpen
  \bibfield  {author} {\bibinfo {author} {\bibfnamefont {G.}~\bibnamefont
  {Iyengar}}\ and\ \bibinfo {author} {\bibfnamefont {M.}~\bibnamefont {Rao}},\
  }\href {http://www.pnas.org/content/111/34/12402.abstract} {\bibfield
  {journal} {\bibinfo  {journal} {Proceedings of the National Academy of
  Sciences}\ }\textbf {\bibinfo {volume} {111}},\ \bibinfo {pages} {12402}
  (\bibinfo {year} {2014})}\BibitemShut {NoStop}%
\bibitem [{\citenamefont {Selimkhanov}\ \emph {et~al.}(2014)\citenamefont
  {Selimkhanov}, \citenamefont {Taylor}, \citenamefont {Yao}, \citenamefont
  {Pilko}, \citenamefont {Albeck}, \citenamefont {Hoffmann}, \citenamefont
  {Tsimring},\ and\ \citenamefont {Wollman}}]{Selimkhanov2014Accurate}%
  \BibitemOpen
  \bibfield  {author} {\bibinfo {author} {\bibfnamefont {J.}~\bibnamefont
  {Selimkhanov}}, \bibinfo {author} {\bibfnamefont {B.}~\bibnamefont {Taylor}},
  \bibinfo {author} {\bibfnamefont {J.}~\bibnamefont {Yao}}, \bibinfo {author}
  {\bibfnamefont {A.}~\bibnamefont {Pilko}}, \bibinfo {author} {\bibfnamefont
  {J.}~\bibnamefont {Albeck}}, \bibinfo {author} {\bibfnamefont
  {A.}~\bibnamefont {Hoffmann}}, \bibinfo {author} {\bibfnamefont
  {L.}~\bibnamefont {Tsimring}}, \ and\ \bibinfo {author} {\bibfnamefont
  {R.}~\bibnamefont {Wollman}},\ }\href
  {http://www.sciencemag.org/content/346/6215/1370.abstract} {\bibfield
  {journal} {\bibinfo  {journal} {Science}\ }\textbf {\bibinfo {volume}
  {346}},\ \bibinfo {pages} {1370} (\bibinfo {year} {2014})}\BibitemShut
  {NoStop}%
\bibitem [{\citenamefont {Murugan}\ \emph {et~al.}(2014)\citenamefont
  {Murugan}, \citenamefont {Huse},\ and\ \citenamefont
  {Leibler}}]{Murugan2014Discriminatory}%
  \BibitemOpen
  \bibfield  {author} {\bibinfo {author} {\bibfnamefont {A.}~\bibnamefont
  {Murugan}}, \bibinfo {author} {\bibfnamefont {D.~A.}\ \bibnamefont {Huse}}, \
  and\ \bibinfo {author} {\bibfnamefont {S.}~\bibnamefont {Leibler}},\
  }\href@noop {} {\bibfield  {journal} {\bibinfo  {journal} {Physical Review
  X}\ }\textbf {\bibinfo {volume} {4}},\ \bibinfo {pages} {021016} (\bibinfo
  {year} {2014})}\BibitemShut {NoStop}%
\bibitem [{\citenamefont {Barato}\ \emph {et~al.}(2014)\citenamefont {Barato},
  \citenamefont {Hartich},\ and\ \citenamefont
  {Seifert}}]{Barato2014Efficiency}%
  \BibitemOpen
  \bibfield  {author} {\bibinfo {author} {\bibfnamefont {A.~C.}\ \bibnamefont
  {Barato}}, \bibinfo {author} {\bibfnamefont {D.}~\bibnamefont {Hartich}}, \
  and\ \bibinfo {author} {\bibfnamefont {U.}~\bibnamefont {Seifert}},\
  }\href@noop {} {\bibfield  {journal} {\bibinfo  {journal} {New Journal of
  Physics}\ }\textbf {\bibinfo {volume} {16}},\ \bibinfo {pages} {103024 
  1367} (\bibinfo {year} {2014})}\BibitemShut {NoStop}%
\bibitem [{\citenamefont {Ito}\ and\ \citenamefont
  {Sagawa}(2014)}]{Ito2014Maxwells}%
  \BibitemOpen
  \bibfield  {author} {\bibinfo {author} {\bibfnamefont {S.}~\bibnamefont
  {Ito}}\ and\ \bibinfo {author} {\bibfnamefont {T.}~\bibnamefont {Sagawa}},\
  }\href@noop {} {\bibfield  {journal} {\bibinfo  {journal} {arXiv preprint
  arXiv:1406.5810}\ } (\bibinfo {year} {2014})}\BibitemShut {NoStop}%
\bibitem [{\citenamefont {Sartori}\ \emph {et~al.}(2014)\citenamefont
  {Sartori}, \citenamefont {Granger}, \citenamefont {Lee},\ and\ \citenamefont
  {Horowitz}}]{Sartori2014Thermodynamic}%
  \BibitemOpen
  \bibfield  {author} {\bibinfo {author} {\bibfnamefont {P.}~\bibnamefont
  {Sartori}}, \bibinfo {author} {\bibfnamefont {L.}~\bibnamefont {Granger}},
  \bibinfo {author} {\bibfnamefont {C.~F.}\ \bibnamefont {Lee}}, \ and\
  \bibinfo {author} {\bibfnamefont {J.~M.}\ \bibnamefont {Horowitz}},\ }\href
  {http://dx.doi.org/10.1371%2Fjournal.pcbi.1003974} {\bibfield  {journal}
  {\bibinfo  {journal} {PLoS Comput Biol}\ }\textbf {\bibinfo {volume} {10}},\
  \bibinfo {pages} {e1003974 EP } (\bibinfo {year} {2014})}\BibitemShut
  {NoStop}%
\bibitem [{\citenamefont {Ouldridge}\ \emph {et~al.}(2015)\citenamefont
  {Ouldridge}, \citenamefont {Govern},\ and\ \citenamefont
  {Wolde}}]{Ouldridge2015On-the-connection}%
  \BibitemOpen
  \bibfield  {author} {\bibinfo {author} {\bibfnamefont {T.~E.}\ \bibnamefont
  {Ouldridge}}, \bibinfo {author} {\bibfnamefont {C.~C.}\ \bibnamefont
  {Govern}}, \ and\ \bibinfo {author} {\bibfnamefont {P.~R.~t.}\ \bibnamefont
  {Wolde}},\ }\href@noop {} {\bibfield  {journal} {\bibinfo  {journal} {arXiv
  preprint arXiv:1503.00909}\ } (\bibinfo {year} {2015})}\BibitemShut {NoStop}%
\bibitem [{\citenamefont {Hartich}\ \emph {et~al.}(2015)\citenamefont
  {Hartich}, \citenamefont {Barato},\ and\ \citenamefont
  {Seifert}}]{Hartich2015Nonequilibrium}%
  \BibitemOpen
  \bibfield  {author} {\bibinfo {author} {\bibfnamefont {D.}~\bibnamefont
  {Hartich}}, \bibinfo {author} {\bibfnamefont {A.~C.}\ \bibnamefont {Barato}},
  \ and\ \bibinfo {author} {\bibfnamefont {U.}~\bibnamefont {Seifert}},\
  }\href@noop {} {\bibfield  {journal} {\bibinfo  {journal} {arXiv preprint
  arXiv:1502.02594}\ } (\bibinfo {year} {2015})}\BibitemShut {NoStop}%
\bibitem [{\citenamefont {Bennett}(1982)}]{Bennett1982The-thermodynamics}%
  \BibitemOpen
  \bibfield  {author} {\bibinfo {author} {\bibfnamefont {C.~H.}\ \bibnamefont
  {Bennett}},\ }\href@noop {} {\bibfield  {journal} {\bibinfo  {journal}
  {International Journal of Theoretical Physics}\ }\textbf {\bibinfo {volume}
  {21}},\ \bibinfo {pages} {905} (\bibinfo {year} {1982})}\BibitemShut
  {NoStop}%
\bibitem [{\citenamefont {Bennett}(2003)}]{Bennett2003Notes}%
  \BibitemOpen
  \bibfield  {author} {\bibinfo {author} {\bibfnamefont {C.}~\bibnamefont
  {Bennett}},\ }\href@noop {} {\bibfield  {journal} {\bibinfo  {journal}
  {Studies In History and Philosophy of Science Part B: Studies In History and
  Philosophy of Modern Physics}\ }\textbf {\bibinfo {volume} {34}},\ \bibinfo
  {pages} {501} (\bibinfo {year} {2003})}\BibitemShut {NoStop}%
\bibitem [{\citenamefont {Berut}\ \emph {et~al.}(2012)\citenamefont {Berut},
  \citenamefont {Arakelyan}, \citenamefont {Petrosyan}, \citenamefont
  {Ciliberto}, \citenamefont {Dillenschneider},\ and\ \citenamefont
  {Lutz}}]{Berut2012Experimental}%
  \BibitemOpen
  \bibfield  {author} {\bibinfo {author} {\bibfnamefont {A.}~\bibnamefont
  {Berut}}, \bibinfo {author} {\bibfnamefont {A.}~\bibnamefont {Arakelyan}},
  \bibinfo {author} {\bibfnamefont {A.}~\bibnamefont {Petrosyan}}, \bibinfo
  {author} {\bibfnamefont {S.}~\bibnamefont {Ciliberto}}, \bibinfo {author}
  {\bibfnamefont {R.}~\bibnamefont {Dillenschneider}}, \ and\ \bibinfo {author}
  {\bibfnamefont {E.}~\bibnamefont {Lutz}},\ }\href
  {http://dx.doi.org/10.1038/nature10872} {\bibfield  {journal} {\bibinfo
  {journal} {Nature}\ }\textbf {\bibinfo {volume} {483}},\ \bibinfo {pages}
  {187} (\bibinfo {year} {2012})}\BibitemShut {NoStop}%
\bibitem [{\citenamefont {Jun}\ \emph {et~al.}(2014)\citenamefont {Jun},
  \citenamefont {Gavrilov},\ and\ \citenamefont
  {Bechhoefer}}]{Jun2014High-Precision}%
  \BibitemOpen
  \bibfield  {author} {\bibinfo {author} {\bibfnamefont {Y.}~\bibnamefont
  {Jun}}, \bibinfo {author} {\bibfnamefont {M.}~\bibnamefont {Gavrilov}}, \
  and\ \bibinfo {author} {\bibfnamefont {J.}~\bibnamefont {Bechhoefer}},\
  }\href {http://link.aps.org/doi/10.1103/PhysRevLett.113.190601} {\bibfield
  {journal} {\bibinfo  {journal} {Physical Review Letters}\ }\textbf {\bibinfo
  {volume} {113}},\ \bibinfo {pages} {190601} (\bibinfo {year}
  {2014})}\BibitemShut {NoStop}%
\bibitem [{\citenamefont {Vedral}(2014)}]{Vedral2014Quantum}%
  \BibitemOpen
  \bibfield  {author} {\bibinfo {author} {\bibfnamefont {V.}~\bibnamefont
  {Vedral}},\ }\href {http://dx.doi.org/10.1038/nphys2904} {\bibfield
  {journal} {\bibinfo  {journal} {Nat Phys}\ }\textbf {\bibinfo {volume}
  {10}},\ \bibinfo {pages} {256} (\bibinfo {year} {2014})}\BibitemShut
  {NoStop}%
\bibitem [{\citenamefont {Vaikuntanathan}\ and\ \citenamefont
  {Jarzynski}(2011)}]{Vaikuntanathan2011Modeling}%
  \BibitemOpen
  \bibfield  {author} {\bibinfo {author} {\bibfnamefont {S.}~\bibnamefont
  {Vaikuntanathan}}\ and\ \bibinfo {author} {\bibfnamefont {C.}~\bibnamefont
  {Jarzynski}},\ }\href {http://link.aps.org/doi/10.1103/PhysRevE.83.061120}
  {\bibfield  {journal} {\bibinfo  {journal} {Physical Review E}\ }\textbf
  {\bibinfo {volume} {83}},\ \bibinfo {pages} {061120} (\bibinfo {year}
  {2011})}\BibitemShut {NoStop}%
\bibitem [{\citenamefont {Jarzynski}(2011)}]{Jarzynski2011Equalities}%
  \BibitemOpen
  \bibfield  {author} {\bibinfo {author} {\bibfnamefont {C.}~\bibnamefont
  {Jarzynski}},\ }\href {\doibase 10.1146/annurev-conmatphys-062910-140506}
  {\bibfield  {journal} {\bibinfo  {journal} {Annual Review of Condensed Matter
  Physics}\ }\textbf {\bibinfo {volume} {2}},\ \bibinfo {pages} {329} (\bibinfo
  {year} {2011})}\BibitemShut {NoStop}%
\bibitem [{\citenamefont {Mandal}\ and\ \citenamefont
  {Jarzynski}(2012)}]{Mandal2012Work}%
  \BibitemOpen
  \bibfield  {author} {\bibinfo {author} {\bibfnamefont {D.}~\bibnamefont
  {Mandal}}\ and\ \bibinfo {author} {\bibfnamefont {C.}~\bibnamefont
  {Jarzynski}},\ }\href {http://www.pnas.org/content/109/29/11641.abstract}
  {\bibfield  {journal} {\bibinfo  {journal} {Proceedings of the National
  Academy of Sciences}\ }\textbf {\bibinfo {volume} {109}},\ \bibinfo {pages}
  {11641} (\bibinfo {year} {2012})}\BibitemShut {NoStop}%
\bibitem [{\citenamefont {Vaikuntanathan}\ \emph {et~al.}(2014)\citenamefont
  {Vaikuntanathan}, \citenamefont {Gingrich},\ and\ \citenamefont
  {Geissler}}]{Vaikuntanathan2014Dynamic}%
  \BibitemOpen
  \bibfield  {author} {\bibinfo {author} {\bibfnamefont {S.}~\bibnamefont
  {Vaikuntanathan}}, \bibinfo {author} {\bibfnamefont {T.~R.}\ \bibnamefont
  {Gingrich}}, \ and\ \bibinfo {author} {\bibfnamefont {P.~L.}\ \bibnamefont
  {Geissler}},\ }\href {http://link.aps.org/doi/10.1103/PhysRevE.89.062108}
  {\bibfield  {journal} {\bibinfo  {journal} {Physical Review E}\ }\textbf
  {\bibinfo {volume} {89}},\ \bibinfo {pages} {062108} (\bibinfo {year}
  {2014})}\BibitemShut {NoStop}%
\bibitem [{\citenamefont {Diamantini}\ and\ \citenamefont
  {Trugenberger}(2014)}]{Diamantini2014Generalized}%
  \BibitemOpen
  \bibfield  {author} {\bibinfo {author} {\bibfnamefont {M.~C.}\ \bibnamefont
  {Diamantini}}\ and\ \bibinfo {author} {\bibfnamefont {C.~A.}\ \bibnamefont
  {Trugenberger}},\ }\href {http://link.aps.org/doi/10.1103/PhysRevE.89.052138}
  {\bibfield  {journal} {\bibinfo  {journal} {Physical Review E}\ }\textbf
  {\bibinfo {volume} {89}},\ \bibinfo {pages} {052138} (\bibinfo {year}
  {2014})}\BibitemShut {NoStop}%
\bibitem [{\citenamefont {Das}(2014)}]{Das2014Capturing}%
  \BibitemOpen
  \bibfield  {author} {\bibinfo {author} {\bibfnamefont {M.}~\bibnamefont
  {Das}},\ }\href {http://link.aps.org/doi/10.1103/PhysRevE.90.062120}
  {\bibfield  {journal} {\bibinfo  {journal} {Physical Review E}\ }\textbf
  {\bibinfo {volume} {90}},\ \bibinfo {pages} {062120} (\bibinfo {year}
  {2014})}\BibitemShut {NoStop}%
\bibitem [{\citenamefont {Parrondo}\ \emph {et~al.}(2015)\citenamefont
  {Parrondo}, \citenamefont {Horowitz},\ and\ \citenamefont
  {Sagawa}}]{Parrondo2015Thermodynamics}%
  \BibitemOpen
  \bibfield  {author} {\bibinfo {author} {\bibfnamefont {J.~M.~R.}\
  \bibnamefont {Parrondo}}, \bibinfo {author} {\bibfnamefont {J.~M.}\
  \bibnamefont {Horowitz}}, \ and\ \bibinfo {author} {\bibfnamefont
  {T.}~\bibnamefont {Sagawa}},\ }\href {http://dx.doi.org/10.1038/nphys3230}
  {\bibfield  {journal} {\bibinfo  {journal} {Nat Phys}\ }\textbf {\bibinfo
  {volume} {11}},\ \bibinfo {pages} {131} (\bibinfo {year} {2015})}\BibitemShut
  {NoStop}%
\bibitem [{\citenamefont {Hopfield}(1974)}]{Hopfield1974Kinetic}%
  \BibitemOpen
  \bibfield  {author} {\bibinfo {author} {\bibfnamefont {J.~J.}\ \bibnamefont
  {Hopfield}},\ }\href {http://www.pnas.org/content/71/10/4135.abstract}
  {\bibfield  {journal} {\bibinfo  {journal} {Proceedings of the National
  Academy of Sciences}\ }\textbf {\bibinfo {volume} {71}},\ \bibinfo {pages}
  {4135} (\bibinfo {year} {1974})}\BibitemShut {NoStop}%
\bibitem [{\citenamefont {Mckeithan}(1995)}]{Mckeithan1995Kinetic}%
  \BibitemOpen
  \bibfield  {author} {\bibinfo {author} {\bibfnamefont {T.~W.}\ \bibnamefont
  {Mckeithan}},\ }\href@noop {} {\bibfield  {journal} {\bibinfo  {journal}
  {Proceedings of the national academy of sciences}\ }\textbf {\bibinfo
  {volume} {92}},\ \bibinfo {pages} {5042} (\bibinfo {year}
  {1995})}\BibitemShut {NoStop}%
\bibitem [{\citenamefont {Murugan}\ \emph {et~al.}(2012)\citenamefont
  {Murugan}, \citenamefont {Huse},\ and\ \citenamefont
  {Leibler}}]{Murugan2012Speed}%
  \BibitemOpen
  \bibfield  {author} {\bibinfo {author} {\bibfnamefont {A.}~\bibnamefont
  {Murugan}}, \bibinfo {author} {\bibfnamefont {D.~A.}\ \bibnamefont {Huse}}, \
  and\ \bibinfo {author} {\bibfnamefont {S.}~\bibnamefont {Leibler}},\ }\href
  {http://www.pnas.org/content/109/30/12034.abstract} {\bibfield  {journal}
  {\bibinfo  {journal} {Proceedings of the National Academy of Sciences}\
  }\textbf {\bibinfo {volume} {109}},\ \bibinfo {pages} {12034} (\bibinfo
  {year} {2012})}\BibitemShut {NoStop}%
\bibitem [{\citenamefont {Cameron}\ and\ \citenamefont
  {Collins}(2014)}]{Cameron2014Tunable}%
  \BibitemOpen
  \bibfield  {author} {\bibinfo {author} {\bibfnamefont {D.~E.}\ \bibnamefont
  {Cameron}}\ and\ \bibinfo {author} {\bibfnamefont {J.~J.}\ \bibnamefont
  {Collins}},\ }\href {http://dx.doi.org/10.1038/nbt.3053} {\bibfield
  {journal} {\bibinfo  {journal} {Nat Biotech}\ }\textbf {\bibinfo {volume}
  {32}},\ \bibinfo {pages} {1276} (\bibinfo {year} {2014})}\BibitemShut
  {NoStop}%
\bibitem [{\citenamefont {Del~Vecchio}\ \emph {et~al.}(2008)\citenamefont
  {Del~Vecchio}, \citenamefont {Ninfa},\ and\ \citenamefont
  {Sontag}}]{Del-Vecchio2008Modular}%
  \BibitemOpen
  \bibfield  {author} {\bibinfo {author} {\bibfnamefont {D.}~\bibnamefont
  {Del~Vecchio}}, \bibinfo {author} {\bibfnamefont {A.~J.}\ \bibnamefont
  {Ninfa}}, \ and\ \bibinfo {author} {\bibfnamefont {E.~D.}\ \bibnamefont
  {Sontag}},\ }\href {\doibase 10.1038/msb4100204} {\bibfield  {journal}
  {\bibinfo  {journal} {Molecular Systems Biology}\ }\textbf {\bibinfo {volume}
  {4}} (\bibinfo {year} {2008}),\ 10.1038/msb4100204}\BibitemShut {NoStop}%
\bibitem [{\citenamefont {Barton}\ and\ \citenamefont
  {Sontag}(2013)}]{Barton2013The-energy}%
  \BibitemOpen
  \bibfield  {author} {\bibinfo {author} {\bibfnamefont {J.~P.}\ \bibnamefont
  {Barton}}\ and\ \bibinfo {author} {\bibfnamefont {E.~D.}\ \bibnamefont
  {Sontag}},\ }\href@noop {} {\bibfield  {journal} {\bibinfo  {journal}
  {Biophysical journal}\ }\textbf {\bibinfo {volume} {104}},\ \bibinfo {pages}
  {1380} (\bibinfo {year} {2013})}\BibitemShut {NoStop}%
\bibitem [{\citenamefont {Mishra}\ \emph {et~al.}(2014)\citenamefont {Mishra},
  \citenamefont {Rivera}, \citenamefont {Lin}, \citenamefont {Del~Vecchio},\
  and\ \citenamefont {Weiss}}]{Mishra2014A-load}%
  \BibitemOpen
  \bibfield  {author} {\bibinfo {author} {\bibfnamefont {D.}~\bibnamefont
  {Mishra}}, \bibinfo {author} {\bibfnamefont {P.~M.}\ \bibnamefont {Rivera}},
  \bibinfo {author} {\bibfnamefont {A.}~\bibnamefont {Lin}}, \bibinfo {author}
  {\bibfnamefont {D.}~\bibnamefont {Del~Vecchio}}, \ and\ \bibinfo {author}
  {\bibfnamefont {R.}~\bibnamefont {Weiss}},\ }\href@noop {} {\bibfield
  {journal} {\bibinfo  {journal} {Nature biotechnology}\ }\textbf {\bibinfo
  {volume} {32}},\ \bibinfo {pages} {1268} (\bibinfo {year}
  {2014})}\BibitemShut {NoStop}%
\bibitem [{\citenamefont {Bialek}(2012)}]{Bialek2012Biophysics-Searching}%
  \BibitemOpen
  \bibfield  {author} {\bibinfo {author} {\bibfnamefont {W.}~\bibnamefont
  {Bialek}},\ }\href@noop {} {\emph {\bibinfo {title} {Biophysics: Searching
  for Principles}}}\ (\bibinfo  {publisher} {Princeton University Press},\
  \bibinfo {year} {2012})\BibitemShut {NoStop}%
\bibitem [{\citenamefont {Rieke}\ and\ \citenamefont
  {Baylor}(1996)}]{Rieke1996Molecular}%
  \BibitemOpen
  \bibfield  {author} {\bibinfo {author} {\bibfnamefont {F.}~\bibnamefont
  {Rieke}}\ and\ \bibinfo {author} {\bibfnamefont {D.~A.}\ \bibnamefont
  {Baylor}},\ }\href {\doibase 10.1016/S0006-3495(96)79448-1} {\bibfield
  {journal} {\bibinfo  {journal} {Biophysical Journal}\ }\textbf {\bibinfo
  {volume} {71}},\ \bibinfo {pages} {2553} (\bibinfo {year}
  {1996})}\BibitemShut {NoStop}%
\bibitem [{\citenamefont {Rieke}\ and\ \citenamefont
  {Baylor}(1998)}]{Rieke1998Single-photon}%
  \BibitemOpen
  \bibfield  {author} {\bibinfo {author} {\bibfnamefont {F.}~\bibnamefont
  {Rieke}}\ and\ \bibinfo {author} {\bibfnamefont {D.~A.}\ \bibnamefont
  {Baylor}},\ }\href {http://link.aps.org/doi/10.1103/RevModPhys.70.1027}
  {\bibfield  {journal} {\bibinfo  {journal} {Reviews of Modern Physics}\
  }\textbf {\bibinfo {volume} {70}},\ \bibinfo {pages} {1027} (\bibinfo {year}
  {1998})}\BibitemShut {NoStop}%
\bibitem [{\citenamefont {Doan}\ \emph {et~al.}(2006)\citenamefont {Doan},
  \citenamefont {Mendez}, \citenamefont {Detwiler}, \citenamefont {Chen},\ and\
  \citenamefont {Rieke}}]{Doan2006Multiple}%
  \BibitemOpen
  \bibfield  {author} {\bibinfo {author} {\bibfnamefont {T.}~\bibnamefont
  {Doan}}, \bibinfo {author} {\bibfnamefont {A.}~\bibnamefont {Mendez}},
  \bibinfo {author} {\bibfnamefont {P.~B.}\ \bibnamefont {Detwiler}}, \bibinfo
  {author} {\bibfnamefont {J.}~\bibnamefont {Chen}}, \ and\ \bibinfo {author}
  {\bibfnamefont {F.}~\bibnamefont {Rieke}},\ }\href
  {http://www.sciencemag.org/content/313/5786/530.abstract} {\bibfield
  {journal} {\bibinfo  {journal} {Science}\ }\textbf {\bibinfo {volume}
  {313}},\ \bibinfo {pages} {530} (\bibinfo {year} {2006})}\BibitemShut
  {NoStop}%
\bibitem [{\citenamefont {Munsky}\ \emph {et~al.}(2009)\citenamefont {Munsky},
  \citenamefont {Nemenman},\ and\ \citenamefont {Bel}}]{Munsky2009Specificity}%
  \BibitemOpen
  \bibfield  {author} {\bibinfo {author} {\bibfnamefont {B.}~\bibnamefont
  {Munsky}}, \bibinfo {author} {\bibfnamefont {I.}~\bibnamefont {Nemenman}}, \
  and\ \bibinfo {author} {\bibfnamefont {G.}~\bibnamefont {Bel}},\ }\href
  {http://scitation.aip.org/content/aip/journal/jcp/131/23/10.1063/1.3274803}
  {\bibfield  {journal} {\bibinfo  {journal} {The Journal of Chemical Physics}\
  }\textbf {\bibinfo {volume} {131}},\  (\bibinfo {year} {2009})}\BibitemShut
  {NoStop}%
\bibitem [{\citenamefont {Bel}\ \emph {et~al.}(2010)\citenamefont {Bel},
  \citenamefont {Munsky},\ and\ \citenamefont
  {Nemenman}}]{Bel2010The-simplicity}%
  \BibitemOpen
  \bibfield  {author} {\bibinfo {author} {\bibfnamefont {G.}~\bibnamefont
  {Bel}}, \bibinfo {author} {\bibfnamefont {B.}~\bibnamefont {Munsky}}, \ and\
  \bibinfo {author} {\bibfnamefont {I.}~\bibnamefont {Nemenman}},\ }\href
  {http://stacks.iop.org/1478-3975/7/i=1/a=016003} {\bibfield  {journal}
  {\bibinfo  {journal} {Physical Biology}\ }\textbf {\bibinfo {volume} {7}},\
  \bibinfo {pages} {016003} (\bibinfo {year} {2010})}\BibitemShut {NoStop}%
\bibitem [{\citenamefont {Lang}\ \emph {et~al.}(2014)\citenamefont {Lang},
  \citenamefont {Fisher}, \citenamefont {Mora},\ and\ \citenamefont
  {Mehta}}]{Lang2014Thermodynamics}%
  \BibitemOpen
  \bibfield  {author} {\bibinfo {author} {\bibfnamefont {A.~H.}\ \bibnamefont
  {Lang}}, \bibinfo {author} {\bibfnamefont {C.~K.}\ \bibnamefont {Fisher}},
  \bibinfo {author} {\bibfnamefont {T.}~\bibnamefont {Mora}}, \ and\ \bibinfo
  {author} {\bibfnamefont {P.}~\bibnamefont {Mehta}},\ }\href
  {http://link.aps.org/doi/10.1103/PhysRevLett.113.148103} {\bibfield
  {journal} {\bibinfo  {journal} {Physical Review Letters}\ }\textbf {\bibinfo
  {volume} {113}},\ \bibinfo {pages} {148103} (\bibinfo {year}
  {2014})}\BibitemShut {NoStop}%
\bibitem [{\citenamefont {Barato}\ and\ \citenamefont
  {Seifert}(2015)}]{Barato2015Thermodynamic}%
  \BibitemOpen
  \bibfield  {author} {\bibinfo {author} {\bibfnamefont {A.~C.}\ \bibnamefont
  {Barato}}\ and\ \bibinfo {author} {\bibfnamefont {U.}~\bibnamefont
  {Seifert}},\ }\href {http://link.aps.org/doi/10.1103/PhysRevLett.114.158101}
  {\bibfield  {journal} {\bibinfo  {journal} {Physical Review Letters}\
  }\textbf {\bibinfo {volume} {114}},\ \bibinfo {pages} {158101} (\bibinfo
  {year} {2015})}\BibitemShut {NoStop}%
\bibitem [{\citenamefont {Detwiler}\ \emph {et~al.}(2000)\citenamefont
  {Detwiler}, \citenamefont {Ramanathan}, \citenamefont {Sengupta},\ and\
  \citenamefont {Shraiman}}]{Detwiler2000Engineering}%
  \BibitemOpen
  \bibfield  {author} {\bibinfo {author} {\bibfnamefont {P.~B.}\ \bibnamefont
  {Detwiler}}, \bibinfo {author} {\bibfnamefont {S.}~\bibnamefont
  {Ramanathan}}, \bibinfo {author} {\bibfnamefont {A.}~\bibnamefont
  {Sengupta}}, \ and\ \bibinfo {author} {\bibfnamefont {B.~I.}\ \bibnamefont
  {Shraiman}},\ }\href {\doibase
  http://dx.doi.org/10.1016/S0006-3495(00)76519-2} {\bibfield  {journal}
  {\bibinfo  {journal} {Biophysical Journal}\ }\textbf {\bibinfo {volume}
  {79}},\ \bibinfo {pages} {2801} (\bibinfo {year} {2000})}\BibitemShut
  {NoStop}%
\bibitem [{\citenamefont {Berg}\ and\ \citenamefont
  {Purcell}(1977)}]{Berg1977Physics}%
  \BibitemOpen
  \bibfield  {author} {\bibinfo {author} {\bibfnamefont {H.~C.}\ \bibnamefont
  {Berg}}\ and\ \bibinfo {author} {\bibfnamefont {E.~M.}\ \bibnamefont
  {Purcell}},\ }\href
  {http://www.sciencedirect.com/science/article/pii/S0006349577855446}
  {\bibfield  {journal} {\bibinfo  {journal} {Biophysical Journal}\ }\textbf
  {\bibinfo {volume} {20}},\ \bibinfo {pages} {193} (\bibinfo {year}
  {1977})}\BibitemShut {NoStop}%
\bibitem [{\citenamefont {Feynman}\ \emph {et~al.}(1998)\citenamefont
  {Feynman}, \citenamefont {Hey},\ and\ \citenamefont
  {Allen}}]{Feynman1998Feynman}%
  \BibitemOpen
  \bibfield  {author} {\bibinfo {author} {\bibfnamefont {R.~P.}\ \bibnamefont
  {Feynman}}, \bibinfo {author} {\bibfnamefont {J.}~\bibnamefont {Hey}}, \ and\
  \bibinfo {author} {\bibfnamefont {R.~W.}\ \bibnamefont {Allen}},\ }\href@noop
  {} {\emph {\bibinfo {title} {Feynman lectures on computation}}}\ (\bibinfo
  {publisher} {Addison-Wesley Longman Publishing Co., Inc.},\ \bibinfo {year}
  {1998})\BibitemShut {NoStop}%
\bibitem [{\citenamefont {Mehta}\ and\ \citenamefont
  {Schwab}(2012)}]{Mehta2012Energetic}%
  \BibitemOpen
  \bibfield  {author} {\bibinfo {author} {\bibfnamefont {P.}~\bibnamefont
  {Mehta}}\ and\ \bibinfo {author} {\bibfnamefont {D.~J.}\ \bibnamefont
  {Schwab}},\ }\href {http://www.pnas.org/content/109/44/17978.abstract}
  {\bibfield  {journal} {\bibinfo  {journal} {Proceedings of the National
  Academy of Sciences}\ }\textbf {\bibinfo {volume} {109}},\ \bibinfo {pages}
  {17978} (\bibinfo {year} {2012})}\BibitemShut {NoStop}%
\bibitem [{\citenamefont {Bonnet}\ \emph {et~al.}(2012)\citenamefont {Bonnet},
  \citenamefont {Subsoontorn},\ and\ \citenamefont
  {Endy}}]{Bonnet2012Rewritable}%
  \BibitemOpen
  \bibfield  {author} {\bibinfo {author} {\bibfnamefont {J.}~\bibnamefont
  {Bonnet}}, \bibinfo {author} {\bibfnamefont {P.}~\bibnamefont {Subsoontorn}},
  \ and\ \bibinfo {author} {\bibfnamefont {D.}~\bibnamefont {Endy}},\ }\href
  {http://www.pnas.org/content/109/23/8884.abstract} {\bibfield  {journal}
  {\bibinfo  {journal} {Proceedings of the National Academy of Sciences}\
  }\textbf {\bibinfo {volume} {109}},\ \bibinfo {pages} {8884} (\bibinfo {year}
  {2012})}\BibitemShut {NoStop}%
\bibitem [{\citenamefont {Siuti}\ \emph {et~al.}(2013)\citenamefont {Siuti},
  \citenamefont {Yazbek},\ and\ \citenamefont {Lu}}]{Siuti2013Synthetic}%
  \BibitemOpen
  \bibfield  {author} {\bibinfo {author} {\bibfnamefont {P.}~\bibnamefont
  {Siuti}}, \bibinfo {author} {\bibfnamefont {J.}~\bibnamefont {Yazbek}}, \
  and\ \bibinfo {author} {\bibfnamefont {T.~K.}\ \bibnamefont {Lu}},\
  }\href@noop {} {\bibfield  {journal} {\bibinfo  {journal} {Nature
  biotechnology}\ }\textbf {\bibinfo {volume} {31}},\ \bibinfo {pages} {448}
  (\bibinfo {year} {2013})}\BibitemShut {NoStop}%
\bibitem [{\citenamefont {Farzadfard}\ and\ \citenamefont
  {Lu}(2014)}]{farzadfard2014genomically}%
  \BibitemOpen
  \bibfield  {author} {\bibinfo {author} {\bibfnamefont {F.}~\bibnamefont
  {Farzadfard}}\ and\ \bibinfo {author} {\bibfnamefont {T.~K.}\ \bibnamefont
  {Lu}},\ }\href {http://www.sciencemag.org/content/346/6211/1256272.abstract}
  {\bibfield  {journal} {\bibinfo  {journal} {Science}\ }\textbf {\bibinfo
  {volume} {346}} (\bibinfo {year} {2014})}\BibitemShut {NoStop}%
\bibitem [{\citenamefont {Keung}\ \emph {et~al.}(2014)\citenamefont {Keung},
  \citenamefont {Bashor}, \citenamefont {Kiriakov}, \citenamefont {Collins},\
  and\ \citenamefont {Khalil}}]{keung2014using}%
  \BibitemOpen
  \bibfield  {author} {\bibinfo {author} {\bibfnamefont {A.~J.}\ \bibnamefont
  {Keung}}, \bibinfo {author} {\bibfnamefont {C.~J.}\ \bibnamefont {Bashor}},
  \bibinfo {author} {\bibfnamefont {S.}~\bibnamefont {Kiriakov}}, \bibinfo
  {author} {\bibfnamefont {J.~J.}\ \bibnamefont {Collins}}, \ and\ \bibinfo
  {author} {\bibfnamefont {A.~S.}\ \bibnamefont {Khalil}},\ }\href {\doibase
  http://dx.doi.org/10.1016/j.cell.2014.04.047} {\bibfield  {journal} {\bibinfo
   {journal} {Cell}\ }\textbf {\bibinfo {volume} {158}},\ \bibinfo {pages}
  {110} (\bibinfo {year} {2014})}\BibitemShut {NoStop}%
\bibitem [{\citenamefont {Lim}(2010)}]{Lim2010Designing}%
  \BibitemOpen
  \bibfield  {author} {\bibinfo {author} {\bibfnamefont {W.~A.}\ \bibnamefont
  {Lim}},\ }\href {http://dx.doi.org/10.1038/nrm2904} {\bibfield  {journal}
  {\bibinfo  {journal} {Nat Rev Mol Cell Biol}\ }\textbf {\bibinfo {volume}
  {11}},\ \bibinfo {pages} {393} (\bibinfo {year} {2010})}\BibitemShut
  {NoStop}%
\bibitem [{\citenamefont {Bashor}\ \emph {et~al.}(2008)\citenamefont {Bashor},
  \citenamefont {Helman}, \citenamefont {Yan},\ and\ \citenamefont
  {Lim}}]{Bashor2008Using}%
  \BibitemOpen
  \bibfield  {author} {\bibinfo {author} {\bibfnamefont {C.~J.}\ \bibnamefont
  {Bashor}}, \bibinfo {author} {\bibfnamefont {N.~C.}\ \bibnamefont {Helman}},
  \bibinfo {author} {\bibfnamefont {S.}~\bibnamefont {Yan}}, \ and\ \bibinfo
  {author} {\bibfnamefont {W.~A.}\ \bibnamefont {Lim}},\ }\href
  {http://www.sciencemag.org/content/319/5869/1539.abstract} {\bibfield
  {journal} {\bibinfo  {journal} {Science}\ }\textbf {\bibinfo {volume}
  {319}},\ \bibinfo {pages} {1539} (\bibinfo {year} {2008})}\BibitemShut
  {NoStop}%
\bibitem [{\citenamefont {Wei}\ \emph {et~al.}(2012)\citenamefont {Wei},
  \citenamefont {Wong}, \citenamefont {Park}, \citenamefont {Corcoran},
  \citenamefont {Peisajovich}, \citenamefont {Onuffer}, \citenamefont {Weiss},\
  and\ \citenamefont {Lim}}]{Wei2012Bacterial}%
  \BibitemOpen
  \bibfield  {author} {\bibinfo {author} {\bibfnamefont {P.}~\bibnamefont
  {Wei}}, \bibinfo {author} {\bibfnamefont {W.~W.}\ \bibnamefont {Wong}},
  \bibinfo {author} {\bibfnamefont {J.~S.}\ \bibnamefont {Park}}, \bibinfo
  {author} {\bibfnamefont {E.~E.}\ \bibnamefont {Corcoran}}, \bibinfo {author}
  {\bibfnamefont {S.~G.}\ \bibnamefont {Peisajovich}}, \bibinfo {author}
  {\bibfnamefont {J.~J.}\ \bibnamefont {Onuffer}}, \bibinfo {author}
  {\bibfnamefont {A.}~\bibnamefont {Weiss}}, \ and\ \bibinfo {author}
  {\bibfnamefont {W.~A.}\ \bibnamefont {Lim}},\ }\href
  {http://dx.doi.org/10.1038/nature11259} {\bibfield  {journal} {\bibinfo
  {journal} {Nature}\ }\textbf {\bibinfo {volume} {488}},\ \bibinfo {pages}
  {384} (\bibinfo {year} {2012})}\BibitemShut {NoStop}%
\bibitem [{\citenamefont {Valk}\ \emph {et~al.}(2014)\citenamefont {Valk},
  \citenamefont {Venta}, \citenamefont {{\"O}rd}, \citenamefont {Faustova},
  \citenamefont {K{\~o}ivom{\"a}gi},\ and\ \citenamefont
  {Loog}}]{Valk2014Multistep}%
  \BibitemOpen
  \bibfield  {author} {\bibinfo {author} {\bibfnamefont {E.}~\bibnamefont
  {Valk}}, \bibinfo {author} {\bibfnamefont {R.}~\bibnamefont {Venta}},
  \bibinfo {author} {\bibfnamefont {M.}~\bibnamefont {{\"O}rd}}, \bibinfo
  {author} {\bibfnamefont {I.}~\bibnamefont {Faustova}}, \bibinfo {author}
  {\bibfnamefont {M.}~\bibnamefont {K{\~o}ivom{\"a}gi}}, \ and\ \bibinfo
  {author} {\bibfnamefont {M.}~\bibnamefont {Loog}},\ }\href
  {http://www.molbiolcell.org/content/25/22/3456.abstract} {\bibfield
  {journal} {\bibinfo  {journal} {Molecular Biology of the Cell}\ }\textbf
  {\bibinfo {volume} {25}},\ \bibinfo {pages} {3456} (\bibinfo {year}
  {2014})}\BibitemShut {NoStop}%
\end{thebibliography}
%

\end{document}